\documentclass[aps,prd,a4paper,twocolumn,amsmath,showpacs,superscriptaddress,nofootinbib,preprintnumbers]{revtex4-1}
\pdfoutput=1
\usepackage{amssymb,amsmath,latexsym,mathrsfs}
\usepackage{graphicx,subfigure}
\usepackage{epsfig}
\usepackage{varioref,xr-hyper}
\usepackage{color}
\usepackage{multirow}
\usepackage{array}
\usepackage{hyperref}
\usepackage{wasysym}
\usepackage{color}
\usepackage{float}
\usepackage[utf8]{inputenc}
\usepackage[T1]{fontenc}

\hypersetup{
     colorlinks   = true,
     urlcolor     = blue,
     citecolor    = blue,
     linkcolor    = blue,
     menucolor    = blue,
     anchorcolor  = blue,
     filecolor    = blue
}

\begin{document}

\title{New physics in light of the $H_0$ tension: an alternative view}

\author{Sunny Vagnozzi}
\email{sunny.vagnozzi@ast.cam.ac.uk}
\affiliation{Kavli Institute for Cosmology (KICC) and Institute of Astronomy,\\University of Cambridge, Madingley Road, Cambridge CB3 0HA, United Kingdom}

\date{\today}

\begin{abstract}

The strong discrepancy between local and early-time (inverse distance ladder) estimates of the Hubble constant $H_0$ could be pointing towards new physics beyond the concordance $\Lambda$CDM model. Several attempts to address this tension through new physics rely on extended cosmological models, featuring extra free parameters beyond the 6 $\Lambda$CDM parameters. However, marginalizing over additional parameters has the effect of broadening the uncertainties on the inferred parameters (including $H_0$), and it is often the case that within these models the $H_0$ tension is addressed due to larger uncertainties rather than a genuine shift in the central value of $H_0$. In this paper I consider an alternative viewpoint: what happens if a physical theory is able to \textit{fix} the extra parameters to a specific set of non-standard values? In this case, the degrees of freedom of the model are reduced with respect to the standard case where the extra parameters are free to vary. Focusing on the dark energy equation of state $w$ and the effective number of relativistic species $N_{\rm eff}$, I find that physical theories able to fix $w \approx -1.3$ or $N_{\rm eff} \approx 3.95$ would lead to an estimate of $H_0$ from Cosmic Microwave Background, Baryon Acoustic Oscillations, and Type Ia Supernovae data in \textit{perfect} agreement with the local distance ladder estimate, without broadening the uncertainty on the former. These two non-standard models are, from a model-selection perspective, strongly disfavoured with respect to the baseline $\Lambda$CDM model. However, models that predict $N_{\rm eff} \approx 3.45$ would be able to bring the tension down to $1.5\sigma$ while only being weakly disfavored with respect to $\Lambda$CDM, whereas models that predict $w \approx -1.1$ would be able to bring the tension down to $2\sigma$ (at the cost of the preference for $\Lambda$CDM being definite). Finally, I estimate dimensionless multipliers relating variations in $H_0$ to variations in $w$ and $N_{\rm eff}$, which can be used to swiftly repeat the analysis of this paper in light of future more precise local distance ladder estimates of $H_0$, should the tension persist. As a caveat, these results were obtained from the \textit{2015} Planck data release, but these findings are qualitatively largely unaffected were I to use more recent data.

\end{abstract}

\maketitle

\section{Introduction}
\label{sec:intro}

The Hubble constant $H_0$ measures the current expansion rate of the Universe, and is undoubtedly one of the single most important cosmological observables~\cite{Hubble:1929ig,Freedman:2010xv}. While several methods for estimating $H_0$ exist, two in particular have been widely used in the literature. The first and more direct method relies on a classical distance ladder approach, by combining Cepheid period–luminosity relations with absolute distance measurements to local distance anchors, in turn used to calibrate distances to Type Ia Supernovae (SNeIa) host galaxies in the Hubble flow~\cite{Freedman:2000cf,Riess:2016jrr}. Applying this method to observations from the \textit{Hubble Space Telescope} (HST) has provided one of the most precise estimates of $H_0$ to date, yielding $H_0 = (73.24 \pm 1.74)\,{\rm km}\,{\rm s}^{-1}\,{\rm Mpc}^{-1}$~\cite{Riess:2016jrr}.

Alternatively, it is possible to extrapolate the value of $H_0$ from cosmological observations, notably measurements of temperature and polarization anisotropies in the Cosmic Microwave Background (CMB), in combination with low-redshift probes of the expansion history, such as Baryon Acoustic Oscillation (BAO) or SNeIa distance measurements, which help break the geometrical degeneracy inherent in CMB data alone. However, the positions of the acoustic peaks in the CMB anisotropy power spectrum essentially measure an angular scale resulting from the projection of a physical scale (the sound horizon) at last-scattering. Extracting $H_0$ from CMB measurements requires assuming a model for the expansion history of the Universe both prior to and post last-scattering, making this estimate indirect and model-dependent. The usual approach is to assume an underlying $\Lambda$CDM model, featuring a cold dark matter (DM) component and a dark energy (DE) component in the form of a cosmological constant with equation of state (EoS) $w=-1$, and highly successful at describing a wide variety of low- and high-redshift precision cosmological observations~\cite{Riess:1998cb,Perlmutter:1998np,Dunkley:2010ge,Ade:2013zuv,Ade:2013gez,Story:2014hni,Ade:2015xua,Alam:2016hwk,Troxel:2017xyo,Aghanim:2018eyx}. Under this assumption, measurements of temperature and polarization anisotropies from the \textit{Planck} satellite 2015 data release point towards a value of $H_0 = (67.27 \pm 0.66)\,{\rm km}\,{\rm s}^{-1}\,{\rm Mpc}^{-1}$~\cite{Ade:2015xua}.

The two independent estimates of $H_0$ are in tension with each other at a level $>3\sigma$, making this so-called ``$H_0$ tension'' a fascinating problem for modern cosmology~\cite{DiValentino:2017gzb}. Several attempts to address this tension have been pursued in the literature, although no convincing resolution has been found to date. On the one hand, it is possible that either or both the \textit{Planck} and \textit{HST} measurements might suffer from systematics which have not been accounted for. Possible systematics in the \textit{Planck} datasets have been studied in e.g.~\cite{Spergel:2013rxa,Addison:2015wyg,Aghanim:2016sns,Lattanzi:2016dzq,Huang:2018xle}, whereas analogous studies in the case of the \textit{HST} measurements have been conducted for instance in~\cite{Rigault:2013gux,Rigault:2014kaa,Scolnic:2017caz,Jones:2018vbn,Rigault:2018ffm}: in both cases, no obvious solution to the dilemma has been found (see also~\cite{Marra:2013rba,Keenan:2013mfa,Wojtak:2013gda,Odderskov:2014hqa,Romano:2016utn,Fleury:2016fda,Wu:2017fpr,Munoz:2017cll,Hoscheit:2018nfl,Camarena:2018nbr,Shanks:2018rka,Bengaly:2018xko,Kenworthy:2019qwq,Vallejo-Pena:2019agp,Martinelli:2019krf,DiValentino:2020evt}). It is worth pointing out the local measurements of $H_0$ alternative to those of~\cite{Riess:2016jrr} exist in the literature. However, most of these alternative measurements seem to consistently point towards values of $H_0$ significantly higher than the CMB estimate~\cite{Efstathiou:2013via,Cardona:2016ems,Feeney:2017sgx,Dhawan:2017ywl,Follin:2017ljs,Gomez-Valent:2018hwc,Birrer:2018vtm,Burns:2018ggj,Collett:2019hrr,Camarena:2019moy,Wong:2019kwg,Freedman:2019abc} (see however the recent study of~\cite{Lombriser:2019ahl}). Future prospects of measuring $H_0$ from gravitational wave standard sirens are bright and could help arbitrating the tension~\cite{Feeney:2018mkj}.~\footnote{At the time this project was initiated, the estimates in~\cite{Riess:2016jrr} and~\cite{Ade:2015xua} were the most up-to-date local distance ladder and high-redshift measurements respectively. Subsequently, more up-to-date measurements have appeared on both sides (see e.g.~\cite{Riess:2019cxk,Aghanim:2018eyx}), which have actually resulted in the significance of the tension exceeding the $4\sigma$ level. However, at the time this work appeared on arXiv, the  2018 \textit{Planck} likelihood had yet to be released (it was publicly released concurrently with~\cite{Aghanim:2019ame} 2 weeks after this work appeared on arXiv). In any case, the results of this work and the significance of the proposed approach would not change substantially if I were to use the more updated measurements in~\cite{Riess:2019cxk,Aghanim:2018eyx}. Moreover, I have provided simple tools to estimate how much my results would change should one wish to take more updated local measurements of $H_0$ into account. These tools, to be discussed in Sec.~\ref{sec:results}, come in the form of dimensionless multipliers relating variations in $H_0$ to variations in other parameters, see in particular Eqs.~(\ref{eq:mw},\ref{eq:mn}).}

Another perhaps more exciting possibility, widely pursued in recent years, is that the $H_0$ tension might be a sign of physics beyond the $\Lambda$CDM model. Some among the simplest possibilities in this sense involve invoking a phantom DE component (with EoS $w<-1$) or the presence of extra relativistic species in the early Universe (so that $N_{\rm eff}>3.046$, where $N_{\rm eff}$ is the effective number of relativistic species)~\cite{DiValentino:2016hlg,Bernal:2016gxb}. Several other possibilities have been considered in the literature, including but not limited to interactions between DM and DE, interactions between DM and some form of dark radiation, decaying DM, modified gravity, or an early DE component: see e.g.~\cite{Berezhiani:2015yta,Qing-Guo:2016ykt,Tram:2016rcw,Ko:2016uft,Karwal:2016vyq,Kumar:2016zpg,Xia:2016vnp,Ko:2016fcd,Chacko:2016kgg,Prilepina:2016rlq,Zhao:2017cud,Vagnozzi:2017ovm,Kumar:2017dnp,Agrawal:2017rvu,Benetti:2017gvm,Feng:2017nss,Zhao:2017urm,DiValentino:2017zyq,Gariazzo:2017pzb,Dirian:2017pwp,DiValentino:2017iww,Sola:2017znb,Feng:2017mfs,Renk:2017rzu,Yang:2017alx,Buen-Abad:2017gxg,Yang:2017ccc,Raveri:2017jto,DiValentino:2017rcr,DiValentino:2017oaw,Khosravi:2017hfi,Santos:2017alg,Peirone:2017vcq,Benetti:2017juy,Binder:2017lkj,Bolejko:2017fos,Feng:2017usu,Mortsell:2018mfj,Vagnozzi:2018jhn,Nunes:2018xbm,Poulin:2018zxs,Bringmann:2018jpr,Kumar:2018yhh,Yang:2018ubt,Yang:2018euj,Yang:2018xlt,Banihashemi:2018oxo,DEramo:2018vss,Guo:2018ans,Graef:2018fzu,Yang:2018uae,El-Zant:2018bsc,Benevento:2018xcu,Yang:2018qmz,Yang:2018xah,Banihashemi:2018has,Aylor:2018drw,Poulin:2018cxd,Carneiro:2018xwq,Kreisch:2019yzn,Pandey:2019plg,Martinelli:2019dau,Kumar:2019wfs,Vattis:2019efj,Pan:2019jqh,Agrawal:2019lmo,Li:2019san,Yang:2019jwn,Adhikari:2019fvb,Keeley:2019esp,Peirone:2019yjs,Lin:2019qug,Yang:2019qza,Cerdonio:2019oqr,Agrawal:2019dlm,Li:2019ypi,Gelmini:2019deq,Rossi:2019lgt,DiValentino:2019exe,Yang:2019uzo,Archidiacono:2019wdp,Kazantzidis:2019dvk,Desmond:2019ygn,Yang:2019nhz,Nesseris:2019fwr,Pan:2019gop,Visinelli:2019qqu,Cai:2019bdh,Ikeda:2019ckp,Schoneberg:2019wmt,Pan:2019hac,DiValentino:2019dzu,Xiao:2019ccl,Panpanich:2019fxq,Knox:2019rjx,Sola:2019jek,Escudero:2019gvw,DiValentino:2019ffd,Benetti:2019lxu,Yan:2019gbw,Banerjee:2019kgu,Yang:2019uog,DiValentino:2019jae,Cheng:2019bkh,Sakstein:2019fmf,Liu:2019awo,Anchordoqui:2019amx,Hart:2019dxi,Frusciante:2019puu,Akarsu:2019hmw,Ding:2019mmw,Ye:2020btb,Yang:2020zuk,Perez:2020cwa,Pan:2020mst,Choi:2020tqp,Pan:2020bur,Sharov:2020bnk,Krishnan:2020obg,Lucca:2020zjb,DAgostino:2020dhv,Nunes:2020hzy,Hogg:2020rdp,Heinesen:2020sre,Benevento:2020fev,Barker:2020gcp,Dai:2020rfo,Zumalacarregui:2020cjh,Hill:2020osr,Desmond:2020wep,Benisty:2020nql,Gomez-Valent:2020mqn,Akarsu:2020yqa,Ballesteros:2020sik,Haridasu:2020xaa,Alestas:2020mvb,Jedamzik:2020krr,Braglia:2020iik,Chudaykin:2020acu,Izaurieta:2020xpk,Jimenez:2020bgw,Ballardini:2020iws,Anchordoqui:2020znj,Aljaf:2020eqh,Ivanov:2020mfr,DiValentino:2020naf,Braglia:2020bym,Banerjee:2020xcn,Elizalde:2020mfs,Sola:2020lba} for what is inevitably an incomplete list of works examining these possibilities. Despite these considerable efforts, a compelling solution to the $H_0$ tension remains yet to be found (see e.g. the discussion in~\cite{Knox:2019rjx}).

Several attempts to address the $H_0$ tension rely on \textit{extended} cosmological models, e.g.~\cite{DiValentino:2016hlg,Bernal:2016gxb}. In other words, models where additional parameters whose values are usually fixed within the concordance $\Lambda$CDM model are instead allowed to freely vary. Examples of such parameters include the DE EoS $w$, the effective number of relativistic species $N_{\rm eff}$, the curvature density parameter $\Omega_k$, the sum of the neutrino masses $M_{\nu}$, and so on (see~\cite{DiValentino:2015ola} for a full exploration of extended models with up to 12 free parameters). However, allowing additional parameters to vary most often results in larger uncertainties on the inferred cosmological parameters, including $H_0$. The reason is that marginalizing over additional parameters inevitably broadens the posterior of all cosmological parameters, particularly if the latter are strongly correlated/degenerate with the additional parameters. As a result, within extended models oftentimes the Hubble tension is relaxed mostly because of an increase in the uncertainty on $H_0$ as inferred from CMB data, and not due to a genuine shift in the central value.~\footnote{See for instance Page~4 of the slides from the talk by Silvia Galli at the ``Advances in Theoretical Cosmology in Light of Data'' Nordita program, available at \href{http://cosmo-nordita.fysik.su.se/talks/w3/d2/Galli_nordita.pdf}{cosmo-nordita.fysik.su.se/talks/w3/d2/Galli\_nordita.pdf}, where this point is made strongly.}

While not a problem inherent to extended models themselves, it is also worth remarking that several works examining solutions to the $H_0$ tension through extended models do not explore whether the increased model complexity invoked to address the tension is actually warranted by a sufficient improvement in the fit to data. From a statistical perspective, this question can be addressed performing model comparison by computing the Bayesian evidence for the extended models and comparing it to that of a reference model (for instance $\Lambda$CDM). Computing the Bayesian evidence has notoriously been computationally expensive, which is why several works have fallen back on more simplistic model comparison metrics (such as the Akaike or Bayesian information criteria). One final drawback is that many works attempting to address the $H_0$ tension in extended models simply combine high-redshift measurements (such as the CMB, BAO and SNeIa observations) with the local distance ladder $H_0$ measurement (usually in the form of a Gaussian prior), and estimate the inferred values for the additional free parameters. However, without prior knowledge of whether the high-redshift and local distance ladder measurements are consistent within the given model to begin with, such an operation is certainly questionable, if not dangerous altogether.

From the above discussion it is clear that, while attempts to address the $H_0$ tension through extended models certainly have many virtues, they do not come flawless. In view of these issues, it is my goal in this work to approach the question from a different angle, and put forward an alternative way of thinking of new physics solutions to the $H_0$ tension. For simplicity, I will focus on new physics in the form of either phantom dark energy ($w<-1$) or extra relativistic species ($N_{\rm eff}>3.046$). What happens if a physical theory is able to \textit{fix} (or approximately fix) $w$ or $N_{\rm eff}$ to a specific set of non-standard values? In this case, the degrees of freedom of the model are reduced with respect to the standard case where $w$ and $N_{\rm eff}$ are free to vary, and in fact the resulting model would have the same number of degrees of freedom as $\Lambda$CDM. The question I then aim to address is the following: what value of $w$ or $N_{\rm eff}$ would such a physical theory have to predict in order for the high-redshift estimate of $H_0$ from CMB, BAO, and SNeIa data to \textit{perfectly} match the local distance ladder estimate, \textit{i.e.} in order to formally reduce the $H_0$ tension to $\approx 0\sigma$?

As I find in Sec.~\ref{subsec:nonstandard}, the answer to the above question is that such a physical theory should predict $w \approx -1.3$ or $N_{\rm eff} \approx 3.95$.~\footnote{See also~\cite{Nesseris:2019fwr} where a lower but still fixed value was invoked to solve the $H_0$ tension, within an approach similar to the one I am following, and also~\cite{Alestas:2020mvb} for another similar approach focused on the DE EoS $w$.} Note that this approach is very different from the standard one. Within the latter, I would vary either or both $w$ and $N_{\rm eff}$, combine high-redshift CMB data with the local $H_0$ measurement, verify that the $H_0$ tension decreases in significance (possibly due to enlarged uncertainties), and finally infer $w$ and $N_{\rm eff}$ from this dataset combination.

Admittedly, the approach I am following is rather unorthodox, and is in some way a hybrid frequentist-Bayesian approach. However, it does not come without virtues. Most importantly, the main recipients of my results are model-builders, to whom I am providing non-standard parameter values to test against. In fact, provided a physical theory is able to fix $w \approx -1.3$ or $N_{\rm eff} \approx 3.95$, such a physical theory would be \textit{guaranteed} to lead to an estimate of $H_0$ from CMB, BAO, and SNeIa data in perfect agreement with the local distance ladder measurement, possibly balancing reduced tension with quality of fit, as I will discuss in Sec.~\ref{subsec:discussion}. I find it worth clarifying that the existence or not of such physical theories at the time of writing does not in principle undermine the motivation for this work: rather, addressing the question I raised can prompt further model-building activity aimed to test against these parameter values. Nonetheless, as I will show in Sec.~\ref{subsec:models}, there already exist physical models which are able to fix, or approximately fix, $w$ and $N_{\rm eff}$ near their ``sweet spot'' values. The existence of such models at the time of writing further reinforces the motivation behind this work.

Moving ahead to other virtues of the proposed approach, because these alleged physical theories would be able to fix $w$ and $N_{\rm eff}$ which thus do not get marginalized over, the uncertainty on $H_0$ will not increase significantly (if at all) with respect to the same value within $\Lambda$CDM. In other words, if a solution to the $H_0$ tension within such non-standard models/physical theories is found, it will be due to a genuine shift in the central value of $H_0$, and not to a larger error bar. Finally, the fact that the number of free parameters in these physical theories remains the same as in $\Lambda$CDM might play in favour of the models themselves when computing the Bayesian evidence: the latter is in fact generally known to disfavour models with extra parameters, unless the improvement in fit is substantial enough to warrant the addition of the extra free parameters~\cite{Trotta:2008qt,Verde:2009tu,Mortlock:2015iqa,Trotta:2017wnx,Kerscher:2019pzk}.

On the matter of model comparison, one can on very general grounds expect that in order for a model able to fix $w$ or $N_{\rm eff}$ to non-standard values which address the $H_0$ tension, such non-standard values would have to be quite far from the standard $w=-1$ and $N_{\rm eff}=3.046$. It is then interesting to additionally address the following two questions: is there a ``sweet spot'' between decrease in Bayesian evidence and reduction in the $H_0$ tension, \textit{i.e.} are there values of $w$ and $N_{\rm eff}$ which, if predicted/fixed by a physical theory, will lead to a satisfactory reduction in the $H_0$ tension while at the same time not leading to a model which is strongly disfavoured against $\Lambda$CDM? The answer is in fact yes, as I will show in Sec.~\ref{subsec:nonstandard}: the curious reader might want to have a look at Fig.~\ref{fig:H0_w_tension_and_evidence} and Fig.~\ref{fig:H0_neff_tension_and_evidence} for a visual representation of my findings, and to subjectively identify such a sweet spot. The second question I want to address is how the non-standard approach I propose compares to the standard lore of considering extended models. For concreteness, I will compare my approach to the case where $w$ and/or $N_{\rm eff}$ are allowed to freely vary, and show that the non-standard approach of fixing $w$ and $N_{\rm eff}$ actually performs surprisingly better from a statistical point of view. For example, physical theories that predict $N_{\rm eff} \approx 3.45$ would be able to bring the tension down to $1.5\sigma$ while only being weakly disfavored with respect to $\Lambda$CDM, performing as well as the 1-parameter extension of $\Lambda$CDM where $N_{\rm eff}$ is allowed to vary in terms of reduction of tension, but performing better than the latter in terms of Bayesian model comparison (and similarly, although less compellingly, for $w$).

The rest of this paper is then organized as follows. I present theoretical foundations necessary to understand the rest of the work in Sec.~\ref{sec:theory}: in particular, I explain why phantom dark energy or extra radiation can address the $H_0$ tension in Sec.~\ref{subsec:newphysics}, whereas I discuss simple measures to quantify the strength of the $H_0$ tension, before briefly discussing Bayesian evidence and model comparison, in Sec.~\ref{subsec:tension}. I discuss the data and methods used in this work in Sec.~\ref{sec:data}, before presenting my results in Sec.~\ref{sec:results}. In particular, I show the results obtained assuming that a physical theory is able to fix (or approximately fix) $w$ and $N_{\rm eff}$ to non-standard values in Sec.~\ref{subsec:nonstandard}, whereas I show the more standard results obtained allowing $w$ and $N_{\rm eff}$ to vary freely in Sec.~\ref{subsec:extended}, before providing a critical comparison of the two approaches in Sec.~\ref{subsec:discussion}, and discussing examples of the aforementioned physical theories in Sec.~\ref{subsec:models}. Finally, I provide concluding remarks in Sec.~\ref{sec:conclusions}.

\section{Theory}
\label{sec:theory}

Here, I briefly review possible ways of addressing the $H_0$ tension by introducing new physics, with focus on the issue of maintaining the angular scale of the acoustic peaks in the CMB fixed. In particular, I will show why a phantom dark energy component, or extra relativistic species in the early Universe, go in the right direction towards addressing the $H_0$ tension. The reader is invited to consult~\cite{Knox:2019rjx} for a more detailed discussion on these issues, in particular regarding the impact of BAO and SNeIa data on these conclusions. I then move on to briefly discuss measures of tension and aspects of Bayesian model comparison which will be useful in this work.

\subsection{Using new physics to solve the $H_0$ tension}
\label{subsec:newphysics}

Measurements of temperature anisotropies in the CMB have revealed a series of (damped) acoustic peaks. These acoustic peaks constitute the fingerprint of Baryon Acoustic Oscillations (BAOs): sound waves propagating in the baryon-photon plasma prior to photon decoupling, set up by the interplay between gravity and radiation pressure~\cite{Peebles:1970ag,Sunyaev:1970er,Bond:1984fp,Bond:1987ub,Bassett:2009mm}. The first acoustic peak is set up by an oscillation mode which had exactly the time to compress once before freezing as photons decoupled shortly after recombination.

The first acoustic peak of the CMB carries the imprint of the comoving sound horizon at last-scattering $r_s(z_{\star})$, given by the following:
\begin{eqnarray}
r_s(z_{\star}) = \int_{z_{\star}}^{\infty}dz\,\frac{c_s(z)}{H(z)}\,,
\label{eq:rs}
\end{eqnarray}
where $z_{\star} \approx 1100$ denotes the redshift of last-scattering, $H(z)$ denotes the expansion rate, and $c_s(z)$ is the sound speed of the photon-baryon fluid. For most of the expansion history prior to last-scattering, $c_s(z) \approx 1/\sqrt{3}$, before dropping rapidly when matter starts to dominate.

Spatial temperature fluctuations at last-scattering are projected to us as anisotropies on the CMB sky. As a consequence, the first acoustic peak actually carries information on the angular scale $\theta_s$ (usually referred to as the angular scale of the first peak), given by:
\begin{eqnarray}
\theta_s = \frac{r_s(z_{\star})}{D_A(z_{\star})}\,,
\label{eq:thetas}
\end{eqnarray}
where $D_A(z_{\star})$ is the angular diameter distance to the surface of last-scattering, given by:
\begin{eqnarray}
D_A(z_{\star}) = \frac{1}{1+z_{\star}}\int_{0}^{z_{\star}}dz\,\frac{1}{H(z)}\,.
\label{eq:da}
\end{eqnarray}
Measurements of anisotropies in the temperature of the CMB, and in particular the position of the first acoustic peak (which appears at a multipole $\ell \simeq \pi/\theta_s$), accurately fix $\theta_s$. Therefore, any modification to the standard cosmological model aimed at solving the $H_0$ tension should not modify $\theta_s$ in the process.

To make progress, let us express $H(z)$ apppearing in Eqs.~(\ref{eq:rs},\ref{eq:da}) in a more convenient form. Let us first consider for simplicity the $\Lambda$CDM model. At early times, relevant for computing the sound horizon $r_s$ through Eq.~(\ref{eq:rs}), I can express the expansion rate as:
\footnotesize
\begin{eqnarray}
\hskip -1.5 cm \frac{H(z)}{H_0}&\approx& \sqrt{(\Omega_c+\Omega_b)(1+z)^3+\Omega_{\gamma} (1+0.2271N_{\rm eff})(1+z)^4} \nonumber \\
&\propto& \sqrt{(\omega_c+\omega_b)(1+z)^3+\omega_{\gamma}(1+0.2271N_{\rm eff})(1+z)^4}\,,
\label{eq:hzearly}
\end{eqnarray}
\normalsize
where $\Omega_b$, $\Omega_c$, and $\Omega_{\gamma}$ are the density parameters of baryons, cold DM, and photons respectively (the latter essentially fixed by the temperature of the CMB), $N_{\rm eff}$ is the effective number of relativistic species, whose value is fixed to $N_{\rm eff}=3.046$ within the standard cosmological model~\cite{Mangano:2001iu,Mangano:2005cc}.~\footnote{Note that this value was recently updated to $N_{\rm eff}=3.045$ in~\cite{deSalas:2016ztq} and $N_{\rm eff}=3.043$ in~\cite{Bennett:2019ewm}. Since current cosmological data does not have possess the sensitivity required to distinguish a change $\Delta N_{\rm eff} \approx 0.003$, which anyhow does not affect my conclusions regarding $H_0$, I will stick to the standard value $N_{\rm eff}=3.046$, in order to conform to previous literature.} In Eq.~(\ref{eq:hzearly}), I have found it convenient to work with the physical density parameters $\omega_b \equiv \Omega_bh^2$, $\omega_c \equiv \Omega_ch^2$, and $\omega_{\gamma} \equiv \Omega_{\gamma}h^2$ (with $h=H_0/(100\,{\rm km}\,{\rm s}^{-1}\,{\rm Mpc}^{-1})$ the reduced Hubble constant). The reason is that early Universe measurements, and in particular the ratio between the heights of even and odd acoustic peaks in the CMB, as well as the overall height of all peaks, are extremely sensitive to $\omega_b$ and $\omega_c$ rather than $\Omega_b$ and $\Omega_c$. Therefore, when considering the effect of new physics at early times affecting the sound horizon $r_s$, it is convenient to keep $\omega_b$ and $\omega_c$ fixed.

On the other hand, late-time measurements (such as BAO or SNeIa distance measurements) are very sensitive to the density parameter $\Omega_m = \Omega_b+\Omega_c+\Omega_{\nu}$ (which includes the contribution of baryons, cold dark matter, and massive neutrinos), although not at the same level as that to which the CMB is sensitive to $\omega_b$ and $\omega_c$ (in other words, there is more freedom in altering $\Omega_m$ than there is in altering $\omega_b$ and $\omega_c$).~\footnote{In the rest of this work, I will fix the sum of the neutrino masses to $M_{\nu}=0.06\,{\rm eV}$, the minimum mass allowed within the normal ordering, given the current very tight upper limits on $M_{\nu}$~\cite{Palanque-Delabrouille:2015pga,Cuesta:2015iho,Giusarma:2016phn,Gerbino:2016sgw,Vagnozzi:2017ovm,Wang:2017htc,Chen:2017ayg,Nunes:2017xon,Giusarma:2018jei,Choudhury:2018byy,Aghanim:2018eyx,Vagnozzi:2019utt,RoyChoudhury:2019hls,Zhang:2020mox,Li:2020gtk}, which also mildly favour the normal ordering~\cite{Huang:2015wrx,Hannestad:2016fog,Gerbino:2016ehw,Xu:2016ddc,Vagnozzi:2017ovm,Simpson:2017qvj,Schwetz:2017fey,Gariazzo:2018pei,deSalas:2018bym,Mahony:2019fyb,RoyChoudhury:2019hls}. Allowing the neutrino mass to vary would not affect my results significantly.} With this in mind, at late times, relevant for computing the angular diameter distance $D_A(z_{\star})$ through Eq.~(\ref{eq:da}), it is convenient to express the expansion rate as (note that I am implicitly assuming a flat Universe):
\begin{eqnarray}
H(z) \approx H_0\sqrt{\Omega_m(1+z)^3+(1-\Omega_m)}\,.
\label{eq:hzlate}
\end{eqnarray}
The name of the game if I want to address the $H_0$ tension is then the following: I we alter either the early [Eq.~(\ref{eq:hzearly})] or late [Eq.~(\ref{eq:hzlate})] expansion rate in such a way that a value of $H_0$ higher than the one inferred assuming $\Lambda$CDM is now required in order to keep $\theta_s$ [Eq.~(\ref{eq:thetas})] fixed?

One possibility is to lower the sound horizon at last-scattering $r_s(z_{\star})$ in Eq.~(\ref{eq:rs}), by increasing the early-time expansion rate while leaving the late-time expansion rate unchanged. In fact, it is known that the $H_0$ tension can be recast as a mismatch in the sound horizon~\cite{Bernal:2016gxb,Aylor:2018drw}, which should be reduced by $\approx 5\%$ in order to remove the tension. One way of reducing $r_s$ is to raise $N_{\rm eff}$ in Eq.~(\ref{eq:hzearly}) beyond its canonical value of $3.046$.~\footnote{It is worth mentioning well-motivated extensions of the Standard Model of Particle Physics in fact predict forms of dark radiation (for instance hidden photons, sterile neutrinos, thermal axions, and so on)~\cite{Archidiacono:2013fha}, which would raise $N_{\rm eff}$ above its standard value of $3.046$: see e.g.~\cite{Foot:1991bp,Ackerman:2008mha,Kaplan:2009de,Blennow:2012de,CyrRacine:2012fz,Fan:2013yva,Conlon:2013isa,Vogel:2013raa,Petraki:2014uza,Marsh:2014gca,Foot:2014osa,Chacko:2015noa,Agrawal:2016quu,Baum:2016oow,Visinelli:2017imh,Murgia:2017lwo,Visinelli:2017qga,Murgia:2018now,Visinelli:2018utg} for examples of such models.} This has the effect of leaving $D_A(z_{\star})$ unchanged (since the late-time expansion rate is unaffected), which however leads to $\theta_s$ decreasing (since $r_s$ has decreased, but $D_A$ has been left unchanged). To restore $\theta_s$ to its inferred value, I need to increase $H_0$ in Eq.~(\ref{eq:hzlate}), in such a way as to decrease $D_A(z_{\star})$ proportionally to $r_s(z_{\star})$. Therefore, allowing for extra relativistic components in the early Universe and hence raising $N_{\rm eff}$ results in a higher inferred value of $H_0$.

Another possibility, however, is to decrease the late-time expansion rate while leaving the early-time expansion rate unchanged. This operation will leave $r_s(z_{\star})$ unaltered, while increasing $D_A(z_{\star})$. I then need to (re)-decrease $D_A(z_{\star}$) in order to keep $\theta_s$ unchanged, and this can be achieved by increasing $H_0$. How can I decrease the late-time expansion rate without changing $\Omega_m$ (see however~\cite{Colgain:2019pck})? From Eq.~(\ref{eq:hzlate}) it is clear that the only residual freedom consists in altering the DE sector, allowing for a DE component other than a cosmological constant. Let us for simplicity consider a DE component with constant EoS $w \neq -1$. Then, Eq.~(\ref{eq:hzlate}) is modified to (assuming again a flat Universe):
\begin{eqnarray}
H(z) \approx H_0\sqrt{\Omega_m(1+z)^3+(1-\Omega_m)(1+z)^{3(1+w)}}\,,
\label{eq:hzlatew}
\end{eqnarray}
which can also easily be generalized to the case where a dynamical DE component is present, \textit{i.e.} a DE component whose EoS is time-varying. Comparing Eq.~(\ref{eq:hzlate}) and Eq.~(\ref{eq:hzlatew}), it is clear that considering a \textit{phantom} dark energy component (\textit{i.e.} one with $w<-1$) will lower the expansion rate \textit{in the past} with respect to the case where the dark energy is in the form of a cosmological constant.~\footnote{On the contrary in the future, \textit{i.e.} for $z \to -1$, the energy density of a phantom component is larger than that of a cosmological constant with the same density parameter. Note that a more drastic possibility for lowering the late-time expansion rate is to allow for negative energy density in the dark energy sector (a possibility pursued for instance in~\cite{Dutta:2018vmq,Visinelli:2019qqu}). This possibility might actually be motivated from a string theory perspective, given the difficulty faced by attempts to construct stable de Sitter vacua in string theory, whereas stable anti-de Sitter vacua emerge quite naturally~\cite{Vafa:2005ui,Wrase:2010ew,Danielsson:2018ztv,Obied:2018sgi,Palti:2019pca} (see e.g.~\cite{Agrawal:2018own,Achucarro:2018vey,Garg:2018reu,
Kehagias:2018uem,Kinney:2018nny} for cosmological implications of this difficulty). I will not pursue this possibility further here, and restrict to the case where the energy density of the dark energy is positive.} It is worth noting that phantom DE components are generally problematic from a theoretical perspective, as they violate the strong energy condition~\cite{Sawicki:2012pz}. It is generically hard to construct phantom models which are fully under control, although it is possible to construct \textit{effective} phantom components which are theoretically well-behaved (for instance within modified gravity theories or brane-world models), see e.g.~\cite{Sahni:2002dx,Vikman:2004dc,Carroll:2004hc,Das:2005yj,Jhingan:2008ym,Deffayet:2010qz,Pan:2012ki,Nojiri:2013ru,Ludwick:2015dba,Cognola:2016gjy,Sebastiani:2016ras,Ludwick:2017tox,Dutta:2017fjw,Casalino:2018tcd}.

In summary, in this section I have explained how purported solutions to the $H_0$ tension involving new physics should lead to a higher inferred value of $H_0$ without altering the angular scale of the first peak $\theta_s$. Two simple ways for achieving this goal are to increase the expansion rate at early times by increasing $N_{\rm eff}$ (which decreases $r_s(z_{\star})$ while leaving $D_A(z_{\star})$ unchanged, requiring therefore an increase in $H_0$ to keep $\theta_s$ fixed), or to decrease the expansion rate at late times by considering a phantom dark energy component with $w<-1$ (which leaves $r_s(z_{\star})$ unchanged while increasing $D_A(z_{\star})$, requiring therefore an increase in $H_0$ in order to keep $\theta_s$ fixed).

\subsection{Measures of tension and Bayesian evidence}
\label{subsec:tension}

I now move on to discuss simple ways to quantify the degree of tension between different estimates of $H_0$. Let me denote the high-redshift estimate and its uncertainty by $H_0^{\rm cosmo}$ and $\sigma_{\rm cosmo}$ respectively, whereas I denote the local distance ladder estimate and its uncertainty by $H_0^{\rm local}$ and $\sigma_{\rm local}$ respectively. Then, the simplest and most intuitive measure of the degree of tension, used in the majority of works examining the $H_0$ tension, is in terms of number of standard deviations $\#\sigma$, computed as:
\begin{eqnarray}
\#\sigma \equiv \frac{ \left \vert H_0^{\rm cosmo}-H_0^{\rm local} \right \vert }{\sqrt{\sigma_{\rm cosmo}^2+\sigma_{\rm local}^2}}\,.
\label{eq:sigma}
\end{eqnarray}

The $\#\sigma$ measure defined in Eq.~(\ref{eq:sigma}) provides a rather intuitive quantification of the degree of tension between two different inferred values of $H_0$. Furthermore, it is essentially equivalent to the 1D distance measure used in~\cite{Camera:2017vbp,Chang:2018rxd} to examine the $\sigma_8$ tension. Notice however that the $\#\sigma$ measure of tension inherently assumes Gaussian posteriors for both the high-redshift and local distance ladder estimates of $H_0$, and ignores possible tensions in other parameter projections. The $\#\sigma$ measure can also overestimate the tension if strong degeneracies in other parameter dimensions are present.

While the $\#\sigma$ measure is a simple and reasonable zeroth-order measure of tension, there are reasons to prefer alternative measures of tension, as argued in~\cite{Lin:2017ikq}. In particular~\cite{Lin:2017ikq} introduces a so-called index of inconsistency (IOI). Let us consider two datasets denoted by $\boldsymbol{1}$ and $\boldsymbol{2}$. Let us further consider analysing these two datasets within the context of a given model, and inferring mean parameter vectors $\boldsymbol{\mu^{(1)}}$ and $\boldsymbol{\mu^{(2)}}$ and covariance matrices $\boldsymbol{C^{(1)}}$ and $\boldsymbol{C^{(2)}}$ respectively. Then, defining $\boldsymbol{\delta} \equiv \boldsymbol{\mu^{(2)}}-\boldsymbol{\mu^{(1)}}$ and $\boldsymbol{G} \equiv (\boldsymbol{C^{(1)}}+\boldsymbol{C^{(2)}})^{-1}$, the IOI is defined as:
\begin{eqnarray}
{\rm IOI} \equiv \frac{1}{2}\boldsymbol{\delta^T}\boldsymbol{G}\boldsymbol{\delta}\,.
\label{eq:ioi}
\end{eqnarray}
In the work in question, we are actually interested in quantifying the tension in a single parameter, \textit{i.e.} $H_0$. In this case, Eq.~(\ref{eq:ioi}) simplifies considerably and reduces to:
\begin{eqnarray}
{\rm IOI} = \frac{1}{2} \frac{(H_0^{\rm cosmo}-H_0^{\rm local})^2}{\sigma_{\rm cosmo}^2+\sigma_{\rm local}^2}\,.
\label{eq:ioihubble}
\end{eqnarray}
We clearly see that the IOI is closely related to the more intuitive $\#\sigma$ measure defined in Eq.~(\ref{eq:sigma}), with the relation between the two being ${\rm IOI} = (\#\sigma)^2/2$. As argued in~\cite{Lin:2017ikq}, the IOI measures the combined difficulty of each distribution to support/favour the mean of the joint distribution.

In~\cite{Lin:2017ikq}, besides introducing the IOI, the authors also provide an empirical scale (inspired by the Jeffreys' scale, and calibrated to the visual separation between different likelihood contours corresponding to different IOI values) to qualify the degree of tension between two datasets given a certain value of IOI. I report this scale in Tab.~\ref{tab:ioi}.
\begingroup
\begin{center}                   
\begin{table}[!h]                    
\begin{tabular}{|c||c|}                  
\hline
${\rm IOI}$ & Strength of inconsistency \\ \hline \hline
${\rm IOI}<1$ & No significant inconsistency \\ \hline
$1<{\rm IOI}<2.5$ & Weak inconsistency \\ \hline
$2.5<{\rm IOI}<5$ & Moderate inconsistency \\ \hline
${\rm IOI}>5$ & Strong inconsistency \\
\hline
\end{tabular}                                                                               \caption{Scale used to qualitatively interpret the degree of tension between two datasets based on the measured index of inconsistency (IOI), as provided in~\cite{Lin:2017ikq}.}
\label{tab:ioi}
\end{table}                   
\end{center}                   
\endgroup
In this work, I will quantify/qualify the tension between the CMB and local determinations of $H_0$ using both the $\#\sigma$ measure of Eq.~(\ref{eq:sigma}) and the IOI as defined in Eq.~(\ref{eq:ioi}), as well as the scale presented in Tab.~\ref{tab:ioi}. For further discussions on the utility of the IOI as a measure of tension, and advantages compared to other types of measures, I refer the reader to~\cite{Lin:2017ikq}. See also~\cite{Raveri:2015maa,Adhikari:2018wnk,Raveri:2018wln,Kohlinger:2018sxx,Handley:2019wlz,Lin:2019zdn,Garcia-Quintero:2019cgt} for works proposing alternative measures of tension.

Finally, as I discussed in Sec.~\ref{sec:intro}, a significant part of this work will be devoted to computing the Bayesian evidence (with respect to $\Lambda$CDM) of the alternative models I take into consideration for addressing the $H_0$ tension, as encapsulated by the Bayes factor of the alternative model with respect to $\Lambda$CDM. Let us consider a dataset $\mathbf{x}$ and two different models ${\cal M}_i$ and ${\cal M}_j$, described by the parameters $\boldsymbol{\theta}_i$ and $\boldsymbol{\theta}_j$. The two models do not necessarily have to be nested. In fact, in most of the cases I will consider, one model cannot even be recovered as a particular limit of the other. If I assume equal prior probabilities for the two models, the Bayes factor of model ${\cal M}_i$ with respect to model ${\cal M}_j$, $B_{ij}$, is given by:
\begin{eqnarray}
B_{ij} \equiv \frac{\int d\boldsymbol{\theta}_i\, \pi(\boldsymbol{\theta}_i \vert {\cal M}_i) {\cal L}(\mathbf{x} \vert \boldsymbol{\theta}_i,{\cal M}_i)\,,}{\int d\boldsymbol{\theta}_j\, \pi(\boldsymbol{\theta}_j \vert {\cal M}_j) {\cal L}(\mathbf{x} \vert \boldsymbol{\theta}_j,{\cal M}_j)\,,}\,,
\label{eq:bayesfactor}
\end{eqnarray}
where $\pi(\boldsymbol{\theta}_i \vert {\cal M}_i)$ is the prior for the parameters $\boldsymbol{\theta}_i$ and ${\cal L}(\mathbf{x} \vert \boldsymbol{\theta}_i,{\cal M}_i)$ is the likelihood of the data given the model parameters $\boldsymbol{\theta}_i$. A Bayes factor $B_{ij}>1$ (or equivalently $\ln B_{ij}>0$) indicates that model ${\cal M}_i$ is more strongly supported by data than model ${\cal M}_j$.

As with the IOI, the degree of preference corresponding to a certain model with respect to a reference model (usually chosen to be $\Lambda$CDM) can be qualitatively assessed once $\ln B_{ij}$ is computed. The Jeffreys scale is a well known example of scale used for performing a qualitative assessment of model preference based on the value of $\ln B_{ij}$~\cite{Jeffreys:1939xee}. In this work, I will use the revised version of Kass \& Raftery~\cite{Kass:1995loi}, reported in Tab.~\ref{tab:kassraftery}.

\begingroup
\begin{center}
\begin{table}[!h]
\begin{tabular}{|c||c|}
\hline
$\ln B_{ij}$ & Strength of preference for model ${\cal M}_i$ \\ \hline \hline
$0 \leq \ln B_{ij} < 1$ & Weak \\ \hline
$1 \leq \ln B_{ij} < 3$ & Definite \\ \hline
$3 \leq \ln B_{ij} < 5$ & Strong \\ \hline
$\ln B_{ij} \geq 5$ & Very strong \\
\hline
\end{tabular}
\caption{Revised Jeffreys scale used to interpret the values of $\ln B_{ij}$ obtained when comparing two competing models through their Bayesian evidence. A value of $\ln B_{ij}>0$ indicates that model $i$ is favoured with respect to model $j$.}
\label{tab:kassraftery}
\end{table}
\end{center}
\endgroup

Before moving forward, a discussion on Bayesian evidence and Bayes factors is in order. As is clear from Eq.~(\ref{eq:bayesfactor}), the Bayesian evidence and correspondingly Bayes factors (with respect to $\Lambda$CDM) for the extended models where $w$ and $N_{\rm eff}$ are allowed to vary depend strongly on the choice of prior on $w$ and $N_{\rm eff}$ themselves. In this sense, since these priors are somewhat arbitrary, the evidence and Bayes factors themselves are also arbitrary to some degree. Therefore, they should not be over-interpreted, or in any case should be interpreted with great caution. In fact, one can always artificially decrease the evidence for the extended model by ensuring that the prior is large enough so as to cover regions where the likelihood is extremely low. In this sense, it is certainly worth moving towards model comparison tools which depend weakly or do not depend at all on priors, see e.g.~\cite{Gariazzo:2019xhx}. When varying $w$ and $N_{\rm eff}$, I will consider flat priors on these two parameters, with prior edges to be described in the following Section.

\section{Datasets and analysis methodology}
\label{sec:data}

In the following, I described the datasets I use and the methods used to analyze them. I consider a combination of cosmological datasets given by the following:
\begin{itemize}
\item Measurements of Cosmic Microwave Background temperature and polarization anisotropies, as well as their cross-correlations, from the \textit{Planck} 2015 data release~\cite{Ade:2015xua}. In particular, I use a combination of the high-$\ell$ TT likelihood, the low-$\ell$ TT likelihood based on maps recovered with \texttt{Commander}, and polarization data in the low-$\ell$ likelihood. The data is analyzed using the publicly available \textit{Planck} likelihood~\cite{Aghanim:2015xee}. I refer to this dataset as ``CMB'' (note that this dataset is frequently referred to as \textit{PlanckTT}+\textit{lowP} in the literature). Notice that I do not make use of the high-$\ell$ polarization likelihood, as the \textit{Planck} collaboration advises caution on the matter given that their 2015 small-scale polarization measurements might still be contaminated by systematics (such as temperature-polarization leakage).~\footnote{At the time this project was initiated, the 2019 legacy \textit{Planck} likelihood was not yet available. The new likelihood was publicly released in~\cite{Aghanim:2019ame} 2 weeks after this work appeared on arXiv. At any rate, I expect the qualitative and most of the quantitative conclusions reached in this work to be unchanged if I were to use the 2019 legacy \textit{Planck} likelihood, so for simplicity I have chosen not to repeat the analysis using the new likelihood.}
\item Baryon Acoustic Oscillation (BAO) distance measurements from the Six-degree Field Galaxy Survey (6dFGS)~\cite{Beutler:2011hx}, the main galaxy sample of the Sloan Digital Sky Survey Data Release 7 (SDSS-MGS)~\cite{Ross:2014qpa}, and the Baryon Oscillation Spectroscopic Survey Data Release 12 (BOSS DR12)~\cite{Alam:2016hwk}. I refer to this dataset as ``BAO''.
\item Luminosity distance measurements from the Pantheon Supernovae Type-Ia (SNeIa) catalogue~\cite{Scolnic:2017caz}. I refer to this dataset as ``SNe''.
\end{itemize}
The combination of the CMB, BAO, and SNe datasets is referred to as \textit{cosmo}, to reflect the fact that these are cosmological datasets from which $H_0$ can be estimated following an inverse distance ladder approach (see for instance~\cite{Percival:2009xn,Heavens:2014rja,Aubourg:2014yra,Cuesta:2014asa,Verde:2016ccp,Bernal:2016gxb,Abbott:2017smn,Feeney:2018mkj,Lemos:2018smw,Taubenberger:2019qna}), in contrast to the distance ladder approach adopted for the local determination of $H_0$. Using the \textit{cosmo} dataset, I estimate $H_0$ and compare it to the local determination from the \textit{Hubble Space Telescope} (HST) which yields $H_0 = (73.24 \pm 1.74)\,{\rm km}\,{\rm s}^{-1}\,{\rm Mpc}^{-1}$~\cite{Riess:2016jrr}.~\footnote{At the time this project was initiated, the $H_0$ measurement in~\cite{Riess:2016jrr} was among the most recently available ones. Subsequently, more updated local measurements of $H_0$ have become available~\cite{Riess:2019cxk}, which have actually worsened the $H_0$ tension. In any case, the conclusions reached in this work would only be mildly affected if I were to compare the $H_0$ estimate from the \textit{cosmo} dataset with the more updated local measurements in~\cite{Riess:2019cxk}. In any case, I have provided simple tools (to be discussed shortly below) to estimate how much my results would change should one wish to take the measurements of $H_0$ in~\cite{Riess:2019cxk}, or more updated measurements, into account. For more details, see Eqs.~(\ref{eq:mw},\ref{eq:mn}).}

As per my discussion in Sec.~\ref{sec:intro} and Sec.~\ref{sec:theory}, I will envisage the possibility that a physical theory is able to fix selected parameters ($w$ and $N_{\rm eff}$) to non-standard values. In this case, the degrees of freedom of the model are reduced with respect to the standard case where $w$ and $N_{\rm eff}$ are free to vary, and in fact the resulting model would have the same number of degrees of freedom as $\Lambda$CDM. The rationale is that such a physical theory could potentially be preferred (or at least not strongly disfavoured) with respect to the baseline $\Lambda$CDM model from the Bayesian evidence point of view, given that the Bayesian evidence tends to disfavour models with additional free parameters unless the improvement in fit is high enough. This result can prompt model-building activity aimed towards testing against these non-standard values of $w$ and $N_{\rm eff}$ (and I will discuss physical theories which can already achieve this goal at the time of writing in Sec.~\ref{subsec:models}).

With the above considerations in mind, I consider the following models:
\begin{itemize}
\item As baseline model, I consider the concordance $\Lambda$CDM model, described by the usual 6 cosmological parameters: the baryon and cold DM physical density parameters $\omega_b$ and $\omega_c$, the angular size of the sound horizon at last-scattering $\theta_s$, the amplitude and tilt of the primordial power spectrum of scalar fluctuations $A_s$ and $n_s$, and the optical depth to reionization $\tau$. Notice that within this model the DE EoS is fixed to $w=-1$ and the effective number of relativistic species is fixed to $N_{\rm eff}=3.046$.
\item I then consider a class of models, denoted by $\overline{w}$CDM, which are described by 6 free parameters exactly as $\Lambda$CDM but where I assume that a physical theory is able to \textit{fix} $w$ to non-standard values such that $w \neq -1$. Following my earlier discussion in Sec.~\ref{subsec:newphysics}, in order to raise $H_0$ I fix $w$ to non-standard values in the \textit{phantom} regime, \textit{i.e.} the region of parameter space where $w<-1$. Notice that this model is also described by 6 free parameters.
\item I finally consider a class of models, denoted by $\overline{N}\Lambda$CDM, which are described by 6 free parameters exactly as $\Lambda$CDM but where I assume that a physical theory is able to \textit{fix} $N_{\rm eff}$ to non-standard values such that $N_{\rm eff} \neq 3.046$. Following my earlier discussion in Sec.~\ref{subsec:newphysics}, in order to raise $H_0$ I fix $N_{\rm eff}$ to non-standard values such that $N_{\rm eff}>3.046$, \textit{i.e.} I allow for extra relativistic species in the early Universe. Notice that this model is also described by 6 free parameters.
\end{itemize}
In addition, I also wish to compare this non-standard approach to the usual approach wherein extended models (with additional parameters varying) are considered. Therefore, at a later stage I also consider the following extended models:
\begin{itemize}
\item The $\Lambda$CDM+$w$ model, where the equation of state of dark energy $w$ is \textit{varied} in addition to the 6 $\Lambda$CDM parameters. This model is described by 7 free parameters.
\item The $\Lambda$CDM+$N_{\rm eff}$ model, where the effective number of relativistic species $N_{\rm eff}$ is \textit{varied} in addition to the 6 $\Lambda$CDM parameters. This model is described by 7 free parameters.
\item The $\Lambda$CDM+$w$+$N_{\rm eff}$ model, where both the equation of state of dark energy $w$ and the number of relativistic species $N_{\rm eff}$ are \textit{varied} in addition to the 6 $\Lambda$CDM parameters. This model is described by 8 free parameters.
\end{itemize}
For the reader's convenience, I provide a summary of the models considered in this work (along with a full descriptions of their free parameters) in Tab.~\ref{tab:models}. Flat priors have been assumed on all parameters unless otherwise stated. When varying $w$ and/or $N_{\rm eff}$, I adopt flat priors on both parameters, with prior edges given by $[-2;-1/3]$ and $[1;5]$ respectively.

\begin{table*}[!htb]
\begin{tabular}{|c||c|c|c|}
\hline
\textbf{Model} & Free parameters & $\#$ Free parameters & Notes \\
\hline
\hline
$\Lambda$CDM & $\omega_b$, $\omega_c$, $\theta_s$, $A_s$, $n_s$, $\tau$ & 6 & fixed $w=-1$, $N_{\rm eff}=3.046$ \\
\hline
$\overline{w}$CDM & $\omega_b$, $\omega_c$, $\theta_s$, $A_s$, $n_s$, $\tau$ & 6 & fixed $w<-1$, $N_{\rm eff}=3.046$ \\
\hline
$\overline{N}\Lambda$CDM & $\omega_b$, $\omega_c$, $\theta_s$, $A_s$, $n_s$, $\tau$ & 6 & fixed $w=-1$, $N_{\rm eff}>3.046$ \\
\hline
$\Lambda$CDM+$w$ & $\omega_b$, $\omega_c$, $\theta_s$, $A_s$, $n_s$, $\tau$, $w$ & 7 & fixed $N_{\rm eff}=3.046$ \\
\hline
$\Lambda$CDM+$N_{\rm eff}$ & $\omega_b$, $\omega_c$, $\theta_s$, $A_s$, $n_s$, $\tau$, $N_{\rm eff}$ & 7 & fixed $w=-1$ \\
\hline
$\Lambda$CDM+$w$+$N_{\rm eff}$ & $\omega_b$, $\omega_c$, $\theta_s$, $A_s$, $n_s$, $\tau$, $w$, $N_{\rm eff}$ & 8 & None \\
\hline
\end{tabular}
\caption{Summary of the cosmological models considered in this work. Notice that the $\overline{w}$CDM and $\overline{N}\Lambda$CDM are actually classes of models (see text above for more discussions).}
\label{tab:models}
\end{table*}

I sample the posterior distributions of the parameters describing the above models by using Markov Chain Monte Carlo (MCMC) methods. The chains are generated through the cosmological MCMC sampler \texttt{CosmoMC}~\cite{Lewis:2002ah}, and their convergence is monitored through the Gelman-Rubin parameter $R-1$~\cite{Gelman:1992zz}, with $R-1<0.01$ required for the chains to be considered converged. For each of these 6 (classes of) models discussed above, and using the \textit{cosmo} (CMB+BAO+SNe) dataset, I infer the Hubble parameter $H_0$ from the generated MCMC chains (notice that $H_0$ is a derived parameter). I then compare the model-dependent high-redshift estimate of $H_0$ to the local value inferred by HST using the distance ladder approach. I quantify the tension between these two estimates by computing $\#\sigma$ [Eq.~(\ref{eq:sigma})] and the IOI [Eq.~(\ref{eq:ioi})]. The obtained values of IOI are used to qualify the strength of the tension between the two estimates of $H_0$ using the scale in Tab.~\ref{tab:ioi}.

Finally, as discussed in Sec.~\ref{subsec:tension}, I compute the Bayesian evidence to assess whether and to what degree the alternative model I am considering is favoured over the $\Lambda$CDM model. More precisely, I compute the logarithm of the Bayes factor $\ln B_{ij}$, where reference model $j$ is the baseline $\Lambda$CDM model. Therefore, a preference for $\Lambda$CDM will be reflected in a value $\ln B_{ij}<0$. Computing the Bayesian evidence has historically been notoriously computationally expensive. Recently important developments have been reported in~\cite{Heavens:2017afc}, where the possibility of estimating the Bayesian evidence directly from MCMC chains has been considered, resulting in the development of a method which is computationally considerably less expensive than earlier ones.

The method put forward in~\cite{Heavens:2017afc} estimates the Bayesian evidence using $k$th nearest neighbour distances between the MCMC samples, with distances computed using the Mahalanobis distance (which uses the covariance matrix of the parameters as metric). Since nearest neighbour distances depend on the local density of points in parameter space, they allow for the estimation of the overall normalization of the posterior distribution (in other words, the constant relating the number density of MCMC samples to the target density), which is required to estimate the Bayesian evidence. I compute the Bayesian evidence through the method proposed in~\cite{Heavens:2017afc} using the publicly available \texttt{MCEvidence} code.~\footnote{The \texttt{MCEvidence} code is publicly available on \texttt{Github}: \href{https://github.com/yabebalFantaye/MCEvidence}{github.com/yabebalFantaye/MCEvidence}.} The values of $\ln B_{ij}$ I obtain are then used to qualify the strength of the preference for the baseline $\Lambda$CDM model using the modified Jeffreys scale reported in Tab.~\ref{tab:kassraftery}. Alternatively, since the priors on the extra parameters are separable and the baseline $\Lambda$CDM model is nested within the other three extended models I consider, a simple way of computing evidence ratios directly from the MCMC chains would be to use the Savage-Dickey density ratio (SDDR), first introduced in the context of cosmology in~\cite{Trotta:2005ar}. I have checked that the evidence ratios obtained through \texttt{MCEvidence} are in good agreement with those estimated through the SDDR.

\section{Results and discussion}

In the following, I discuss the results obtained using the methods and datasets described in Sec.~\ref{sec:theory} and Sec.~\ref{sec:data}. I begin in Sec.~\ref{subsec:nonstandard} by discussing how the $H_0$ tension is reduced within the non-standard $\overline{w}$CDM and $\overline{N}\Lambda$CDM models, and how much these models are disfavoured compared to $\Lambda$CDM depending on the fixed values of $w$ and $N_{\rm eff}$. I then proceed in Sec.~\ref{subsec:extended} by comparing these results to the more common approach of considering extended models (and in particular the $\Lambda$CDM+$w$, $\Lambda$CDM+$N_{\rm eff}$, and $\Lambda$CDM+$w$+$N_{\rm eff}$ models).
\label{sec:results}

\subsection{Fixing $w$ and $N_{\rm eff}$ to non-standard values}
\label{subsec:nonstandard}

I begin by considering the baseline $\Lambda$CDM model where $w$ and $N_{\rm eff}$ are fixed to their standard values of $-1$ and $3.046$ respectively. Within this model, the high-redshift value of $H_0$ inferred from the \textit{cosmo} (CMB+BAO+SNe) dataset is $H_0 = (67.7 \pm 0.6)\,{\rm km}\,{\rm s}^{-1}\,{\rm Mpc}^{-1}$. Comparing this value to the local distance ladder determination from HST~\cite{Riess:2016jrr}, I find that the level of tension between the two, computed using Eq.~(\ref{eq:sigma}), is $\#\sigma \approx 3.0$. The index of inconsistency, computed using Eq.~(\ref{eq:ioi}), is ${\rm IOI} \approx 4.5$. According to the scale of~\cite{Lin:2017ikq} reported in Tab.~\ref{tab:ioi}, this value indicates a moderate level of inconsistency.

\begin{figure}[!tbh]
\includegraphics[width=1.0\linewidth]{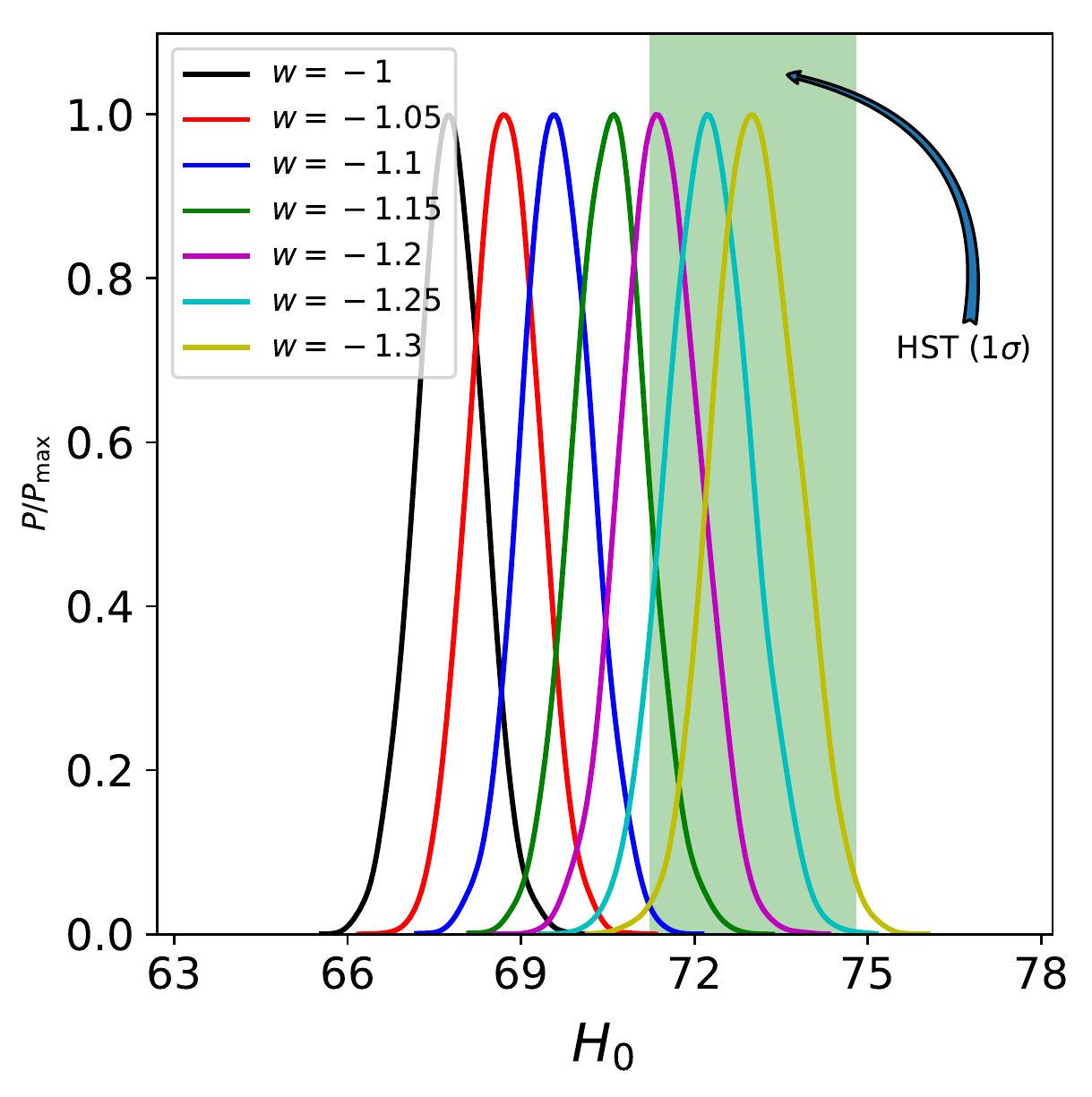}
\caption{Normalized posterior distributions of $H_0$ (in ${\rm km}\,{\rm s}^{-1}\,{\rm Mpc}^{-1}$) for different choices of $w$, where $w$ is the dark energy equation of state fixed to non-standard values within the $\overline{w}$CDM model (see Tab.~\ref{tab:models}). The models considered have values of $w$ fixed to $-1$ (\textit{i.e.} $\Lambda$CDM, black curve), $-1.05$ (red), $-1.1$ (dark blue), $-1.15$ (green), $-1.2$ (purple), $-1.25$ (light blue), and $-1.3$ (yellow). The green shaded region is the 1$\sigma$ credible region for $H_0$ determined by the local distance ladder measurement of \textit{HST}~\cite{Riess:2016jrr}, yielding $H_0 = (73.24 \pm 1.74)\,{\rm km}\,{\rm s}^{-1}\,{\rm Mpc}^{-1}$. When fixing $w=-1.3$, the high-redshift estimate of $H_0$ is $H_0 = (73.2 \pm 0.7)\,{\rm km}\,{\rm s}^{-1}\,{\rm Mpc}^{-1}$, basically in agreement with the local distance ladder measurement.}
\label{fig:H0_posterior_combined_w}
\end{figure}

I then move on to the $\overline{w}$CDM class of models, where a physical theory is assumed to be able to fix the DE EoS $w$ to non-standard values such that $w<-1$ (see Tab.~\ref{tab:models}). The rationale, as explained in Sec.~\ref{sec:theory}, is that one of the simplest possibilities for addressing the $H_0$ tension is by invoking a phantom DE component. For concreteness, I have considered 6 cases where $w$ is fixed to the values $-1.05$, $-1.1$, $-1.15$, $-1.2$, $-1.25$, and $-1.3$ respectively. The normalized posterior distributions for the high-redshift estimate of $H_0$ obtained using the \textit{cosmo} CMB+BAO+SNe dataset combination for these 6 models are shown in Fig.~\ref{fig:H0_posterior_combined_w} (including the $\Lambda$CDM case where $w=-1$). Overlain on the same figure is the 1$\sigma$ region determined by the local distance ladder measurement of \textit{HST}~\cite{Riess:2016jrr}, corresponding to the green shaded area.

I find that if a physical theory were able to fix $w=-1.3$, the high-redshift estimate of $H_0$ inferred from the CMB+BAO+SNe dataset combination is $H_0 = (73.2 \pm 0.7)\,{\rm km}\,{\rm s}^{-1}\,{\rm Mpc}^{-1}$. This value is basically in complete agreement with the local distance ladder estimate of $H_0 = (73.24 \pm 1.74)\,{\rm km}\,{\rm s}^{-1}\,{\rm Mpc}^{-1}$. The level of tension is estimated to be $\#\sigma < 0.1$, with the index of inconsistency being ${\rm IOI} \approx 0$. The uncertainty on the high-redshift estimate of $H_0$ is almost as small as that obtained for the baseline $\Lambda$CDM model. This is expected, given that I am not varying $w$ and hence not marginalizing over it, which would have resulted in a broadening of the constraints on the other parameters. Therefore, barring model comparison considerations (which I will address shortly below), within the $\overline{w}$CDM model with $w \approx -1.3$, the $H_0$ tension is genuinely addressed due to a shift in the mean value of $H_0$, and not due to a significantly larger uncertainty (as often happens in extended models).

Moving on to model comparison considerations, one expects that as $w$ moves away from its standard cosmological constant value $w=-1$, the Bayesian evidence for the corresponding non-standard $\overline{w}$CDM model decreases: in other words, the support for the $\overline{w}$CDM model with respect to $\Lambda$CDM should decrease. To quantify this decrease in support, I have computed $\ln B_{ij}$, with the two competing models being ${\cal M}_i = \overline{w}$CDM and ${\cal M}_j = \Lambda$CDM respectively, and made use of the scale in Tab.~\ref{tab:kassraftery} to interpret the strength of the support for $\Lambda$CDM. Given the definition of $\ln B_{ij}$ and the choice of models $i$ and $j$, a preference for $\Lambda$CDM would be reflected in a value of $\ln B_{ij}<0$.

I find, as expected, that Bayesian evidence model comparison considerations always favour $\Lambda$CDM since $\ln B_{ij}<0$ over all the range of $w$ parameter space considered. In particular, I find that for $-1.07\lesssim w \lesssim-1$, $\Lambda$CDM is weakly preferred over $\overline{w}$CDM, while the preference becomes definite for $-1.14\lesssim w \lesssim -1.07$, strong for $-1.18\lesssim w \lesssim -1.14$, and very strong for $w \lesssim -1.18$. For $w=-1.3$ (which as we saw earlier leads to the high-redshift estimate of $H_0$ agreeing perfectly with the local distance ladder estimate) I find $\ln B_{ij}=-14.9$.

\begin{figure}[!tbh]
\includegraphics[width=1.0\linewidth]{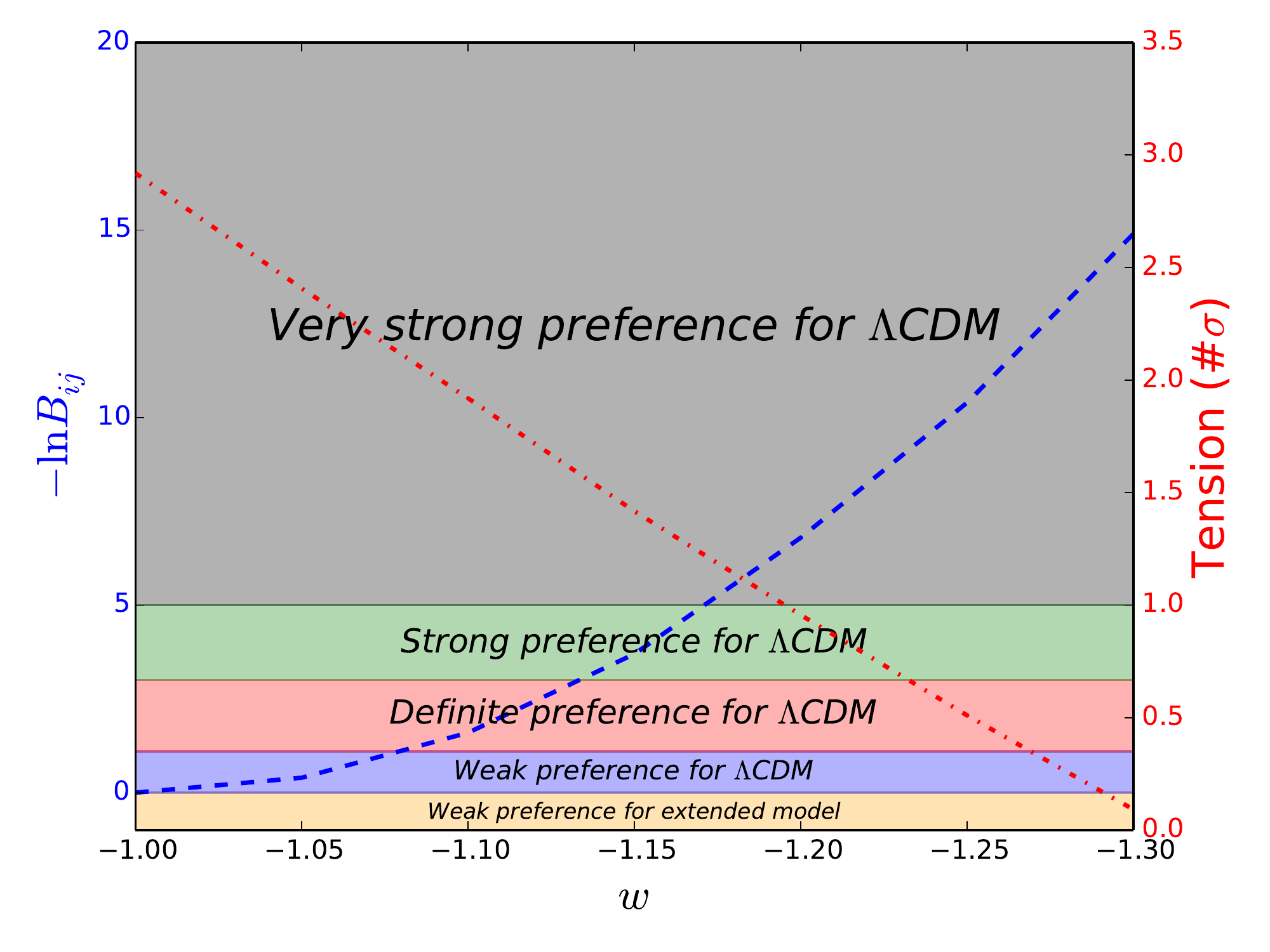}
\caption{Bayesian evidence in favour of $\Lambda$CDM and tension between the high-redshift and local distance latter estimates of $H_0$ as a function of $w$, when the latter is fixed to non-standard values in the phantom region ($w<-1$) within the $\overline{w}$CDM model (see Tab.~\ref{tab:models} for further details). The blue dashed curve (scale on the left $y$-axis) shows $-\ln B_{ij}$ [see Eq.~(\ref{eq:bayesfactor}))], with ${\cal M}_i = \overline{w}$CDM and ${\cal M}_j = \Lambda$CDM. Therefore, a value $-\ln B_{ij}>0$ indicates that $\Lambda$CDM is favoured over the alternative model from the Bayesian evidence point of view. The Jeffreys scale used to quantify the strength of the evidence for $\Lambda$CDM (see Tab.~\ref{tab:kassraftery}) is reflected in the colored regions (orange: weak preference for the extended model; blue: weak preference for $\Lambda$CDM; pink: definite preference for $\Lambda$CDM; green: strong preference for $\Lambda$CDM; grey: very strong preference for $\Lambda$CDM). The red dot-dashed curve quantifies the statistical significance of the $H_0$ tension through $\#\sigma$ [see Eq.~(\ref{eq:sigma})].}
\label{fig:H0_w_tension_and_evidence}
\end{figure}

\begin{figure}[!tbh]
\includegraphics[width=1.0\linewidth]{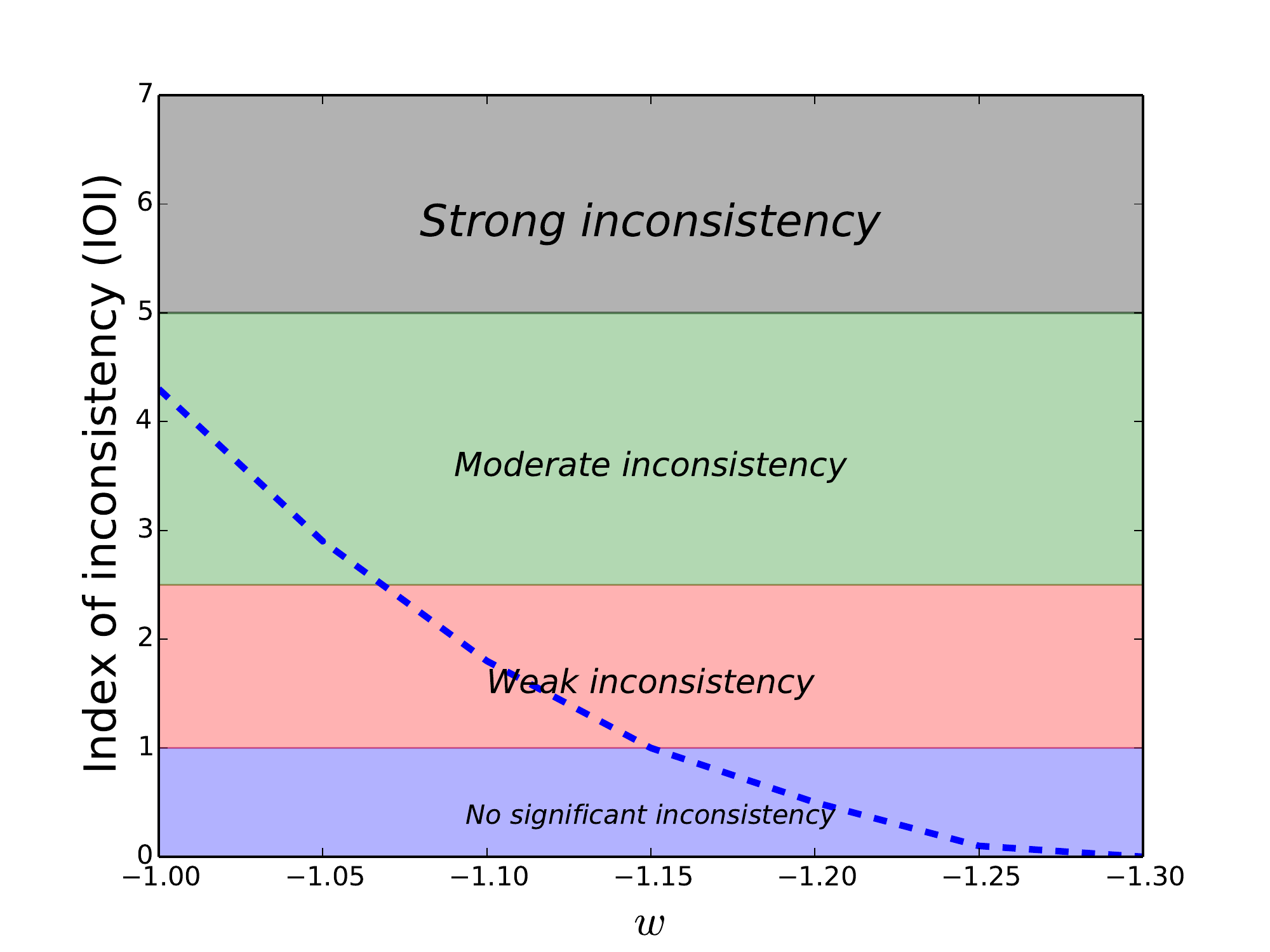}
\caption{Index of inconsistency [see Eq.~(\ref{eq:ioi})] as a function of $w$, when the latter is fixed to non-standard values in the phantom region ($w<-1$) within the $\overline{w}$CDM model (see Tab.~\ref{tab:models} for further details). The scale of~\cite{Lin:2017ikq} used to quantify the strength of the inconsistency is reflected in the colored regions (blue: no significant inconsistency; pink: weak inconsistency; green: moderate inconsistency; grey: strong inconsistency), see Tab.~\ref{tab:ioi} for further details.}
\label{fig:ioi_w}
\end{figure}

A graphical representation of the results discussed so far for the $\overline{w}$CDM model is shown in Fig.~\ref{fig:H0_w_tension_and_evidence} and Fig.~\ref{fig:ioi_w}. In Fig.~\ref{fig:H0_w_tension_and_evidence}, I plot $-\ln B_{ij}$ (left $y$-axis, blue dashed line, note that $-\ln B_{ij}$ which is a positive quantity is being plotted!) and the tension measured in $\#\sigma$ (right $y$-axis, red dot-dashed line) as a function of the fixed value of $w$ in the $\overline{w}$CDM class of models. The figure shows how the tension measured in $\#\sigma$ rapidly decreases as $w$ moves towards $-1.3$, at the cost however of adopting a model which is significantly disfavoured with respect to $\Lambda$CDM (as quantified by the rapidly increasing value of $-\ln B_{ij}$). From the same figure we see that, even accepting a $\overline{w}$CDM model which is weakly disfavoured with respect to $\Lambda$CDM (blue shaded region, $-\ln B_{ij}<1$), the tension cannot be reduced below the $2\sigma$ level. Similarly, even accepting a $\overline{w}$CDM model which is definitely disfavoured with respect to $\Lambda$CDM (pink shaded region, $-\ln B_{ij}<3$), the tension can at best be brought down to the $1.5\sigma$ level (which however some might argue is good enough for the $H_0$ tension to be considered solved).

As one sees from the red dot-dashed curve in Fig.~\ref{fig:H0_w_tension_and_evidence}, as well as the shift in the mean of the posterior distributions in Fig.~\ref{fig:H0_posterior_combined_w}, $H_0$ responds approximately linearly to changes in $w$ when the latter is fixed. In other words, consider a $\overline{w}$CDM model, and define $\Delta w \equiv 1+w$ to be the variation in (the fixed value of) $w$ from the cosmological constant value of $w=-1$. Then, at least for the CMB+BAO+SNe dataset combination, the variation in the central value of $H_0$ from its $\Lambda$CDM value, $\Delta H_0$ (in units of ${\rm km}\,{\rm s}^{-1}\,{\rm Mpc}^{-1}$), should be approximately linearly related to $\Delta w$: $\Delta H_0 \approx m_w\Delta w$, where $m_w$ is a quantity frequently referred to in the literature as dimensionless multiplier, relating variations in different parameters due to a fundamental degeneracy between the two. From my earlier results I numerically estimate $m_w \approx -18.5$, and therefore:
\begin{eqnarray}
\Delta H_0 = H_0-H_0\vert_{\Lambda{\rm CDM}} \approx -18.5\Delta w=-18.5(1+w)\,, \nonumber \\
\label{eq:mw}
\end{eqnarray}
where $H_0\vert_{\Lambda{\rm CDM}}$ is the value of $H_0$ inferred within $\Lambda$CDM. The value $-18.5$ is essentially a reflection of the direction and strength of the $H_0$-$w$ correlation, which I will later show in Fig.~\ref{fig:basew_tri}. The relation in Eq.~(\ref{eq:mw}) is useful especially in light of the fact that local distance ladder measurements of $H_0$ are continuously updated to reflect improvements in analyses techniques. However, Eq.~(\ref{eq:mw}) can always be used to estimate the required fixed value of $w$ to restore agreement with the updated local measurement. For instance, if I were to use the more updated measurement of~\cite{Riess:2019cxk} which yields $H_0=(74.03 \pm 1.42)\,{\rm km}\,{\rm s}^{-1}\,{\rm Mpc}^{-1}$, using Eq.~(\ref{eq:mw}) I would find that a physical theory would need to fix $w \approx -1.35$ in order to restore perfect agreement between the high-redshift and local measurements of $H_0$, \textit{i.e.} a slightly more phantom value compared to what was required for the earlier measurement of~\cite{Riess:2016jrr} which I took as baseline measurement in this work. Of course, the coefficient $-18.5$ in Eq.~(\ref{eq:mw}) is specific for the CMB+BAO+SNe dataset combination, and should eventually be updated if future high-redshift data were to be used, which might change the direction and strength of the correlation in question. One should also keep in mind that the dimensionless multipliers only account for shifts in the central values of the $H_0$ posterior, but do not account for the fact that uncertainties in the local value of $H_0$ are continuously shrinking.

In Fig.~\ref{fig:ioi_w}, I plot the index of inconsistency as a function of the fixed value of $w$ in the $\overline{w}$CDM class of models. One sees that for $-1.07 \lesssim w \lesssim -1$, the inconsistency between the high-redshift and local measurements is moderate, whereas the inconsistency becomes weak for $-1.15 \lesssim w \lesssim -1.07$ and insignificant for $-1.3 \lesssim w \lesssim -1.15$.

\begin{figure}[!tbh]
\includegraphics[width=1.0\linewidth]{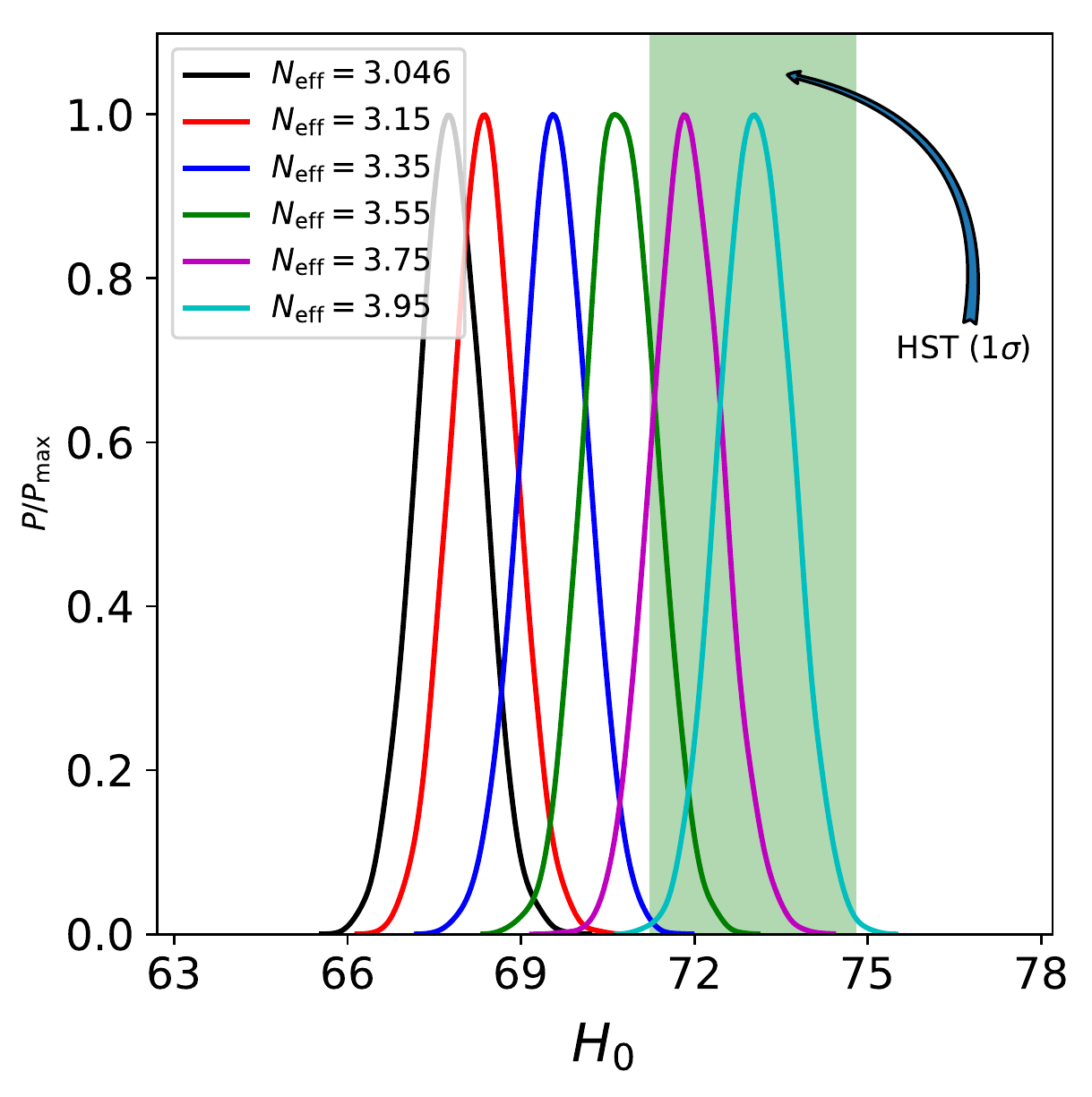}
\caption{Normalized posterior distributions of $H_0$ (in ${\rm km}\,{\rm s}^{-1}\,{\rm Mpc}^{-1}$) for different choices of $N_{\rm eff}$, where $N_{\rm eff}$ is the effective number of relativistic species fixed to non-standard values within the $\overline{N}\Lambda$CDM model (see Tab.~\ref{tab:models}). The models considered have values of $N_{\rm eff}$ fixed to $3.046$ (\textit{i.e.} $\Lambda$CDM, black curve), $3.15$ (red), $3.35$ (dark blue), $3.55$ (green), $3.75$ (purple), and $3.95$ (light blue). The green shaded region is the 1$\sigma$ region of determined by the local distance ladder measurement of HST~\cite{Riess:2016jrr}, yielding $H_0 = (73.24 \pm 1.74)\,{\rm km}\,{\rm s}^{-1}\,{\rm Mpc}^{-1}$. When fixing $N_{\rm eff}=3.95$, the high-redshift estimate of $H_0$ is $H_0 = (73.1 \pm 0.6)\,{\rm km}\,{\rm s}^{-1}\,{\rm Mpc}^{-1}$, in complete agreement with the local distance ladder measurement.}
\label{fig:H0_posterior_combined_neff}
\end{figure}

I now perform a totally analogous analysis within the $\overline{N}\Lambda$CDM model, where a physical theory is assumed to be able to fix the effective number of relativistic species $N_{\rm eff}$ to non-standard values such that $N_{\rm eff}>3.046$ (see Tab.~\ref{tab:models}). The rationale, as explained in Sec.~\ref{sec:theory}, is that the other simple possibility for addressing the $H_0$ tension besides invoking a phantom dark energy component is to allow for extra radiation in the early Universe. For concreteness, I have considered 5 cases where $N_{\rm eff}$ is fixed to the values $3.15$, $3.35$, $3.55$, $3.75$, and $3.95$ respectively. The normalized posterior distributions for the high-redshift estimate of $H_0$ obtained using the \textit{cosmo} CMB+BAO+SNe dataset combination are shown in Fig.~\ref{fig:H0_posterior_combined_neff} (including the $\Lambda$CDM case where $N_{\rm eff}=3.046$). Overlain on the same figure is the 1$\sigma$ region determined by the local distance ladder measurement of HST~\cite{Riess:2016jrr}, corresponding to the green shaded area.

I find that if a physical theory were able to fix $N_{\rm eff}=3.95$, the high-redshift estimate of $H_0$ inferred from the CMB+BAO+SNe dataset combination is $H_0 = (73.1 \pm 0.6)\,{\rm km}\,{\rm s}^{-1}\,{\rm Mpc}^{-1}$. This value is basically in complete agreement with the local distance ladder estimate of $H_0 = (73.24 \pm 1.74)\,{\rm km}\,{\rm s}^{-1}\,{\rm Mpc}^{-1}$. The level of tension is estimated to be $\#\sigma < 0.1$, with the index of inconsistency being ${\rm IOI} \approx 0$. In this case, the uncertainty on the high-redshift estimate of $H_0$ is as small as that obtained for the baseline $\Lambda$CDM model, analogously to what I found assuming that a physical theory were able to fix $w=-1.3$, meaning that the $H_0$ tension is addressed (again barring model comparison considerations to be addressed shortly) due to a genuine shift in the mean value of $H_0$ and not an increase in the error bars.

Analogously to what I did for the $\overline{w}$CDM model, I now compute $\ln B_{ij}$, with the two competing models being ${\cal M}_i = \overline{N}\Lambda$CDM and ${\cal M}_j = \Lambda$CDM respectively, meaning that a preference for $\Lambda$CDM is reflected in a value of $\ln B_{ij}<0$. The results I find are quite similar to those for the $\overline{w}$CDM model, with a small twist. In the range $3.046 \lesssim N_{\rm eff} \lesssim 3.25$, the evidence for the $\overline{N}\Lambda$CDM model is actually \textit{slightly} higher than that of $\Lambda$CDM, and reaches a maximum for $N_{\rm eff} \approx 3.15$, with $\ln B_{ij} \approx 0.2$. Given that according to the scale in Tab.~\ref{tab:kassraftery} a value of $\ln B_{ij} \approx 0.2$ indicates only a weak preference for the $\overline{N}\Lambda$CDM model, I choose not to discuss this feature further.~\footnote{This slight preference for an $\overline{N}\Lambda$CDM model with $N_{\rm eff}=3.15$ might well be due to the fact that \textit{Planck} temperature and large-scale polarization data alone appear to favour a value of $N_{\rm eff}$ slightly higher than the canonical $3.046$ (see~\cite{Ade:2015xua} where $N_{\rm eff}=3.13\pm 0.32$ from the \textit{PlanckTT}+\textit{lowP} dataset combination is reported). This preference disappears when small-scale polarization data is included, especially because small-scale polarization data helps breaking various parameter degeneracies involving $N_{\rm eff}$, and consequently leads to a better determination of this parameter. However, in this work I have made the conservative choice of not including small-scale polarization data, because of possible residual systematics in the 2015 \textit{Planck} dataset (see Sec.~\ref{sec:data}).} For $N_{\rm eff} \gtrsim 3.25$, $\Lambda$CDM is always favoured over the $\overline{N}\Lambda$CDM model from the point of view of Bayesian evidence. In particular, I find that the preference is weak for $3.25 \lesssim N_{\rm eff} \lesssim 3.5$, definite for $3.5 \lesssim N_{\rm eff} \lesssim 3.75$, and strong for $3.75 \lesssim N_{\rm eff} \lesssim 3.9$. For larger values, the preference for $\Lambda$CDM becomes very strong. For $w=3.95$ (which as we saw earlier leads to the high-redshift estimate of $H_0$ agreeing perfectly with the local distance ladder estimate) I find $\ln B_{ij}=-5.5$.

\begin{figure}[!tbh]
\includegraphics[width=1.0\linewidth]{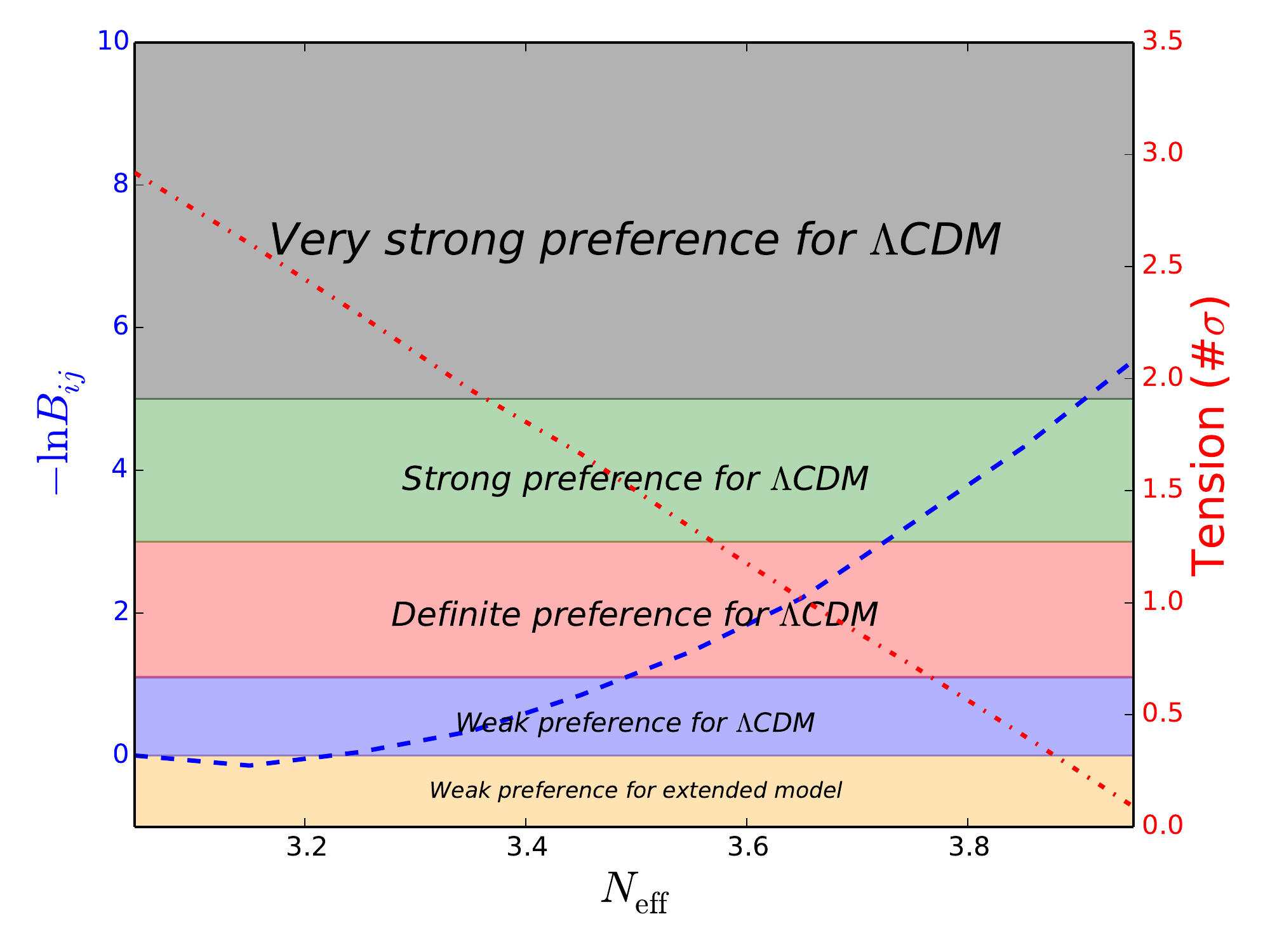}
\caption{As in Fig.~\ref{fig:H0_w_tension_and_evidence} but for the $\overline{N}\Lambda$CDM model.}
\label{fig:H0_neff_tension_and_evidence}
\end{figure}

\begin{figure}[!tbh]
\includegraphics[width=1.0\linewidth]{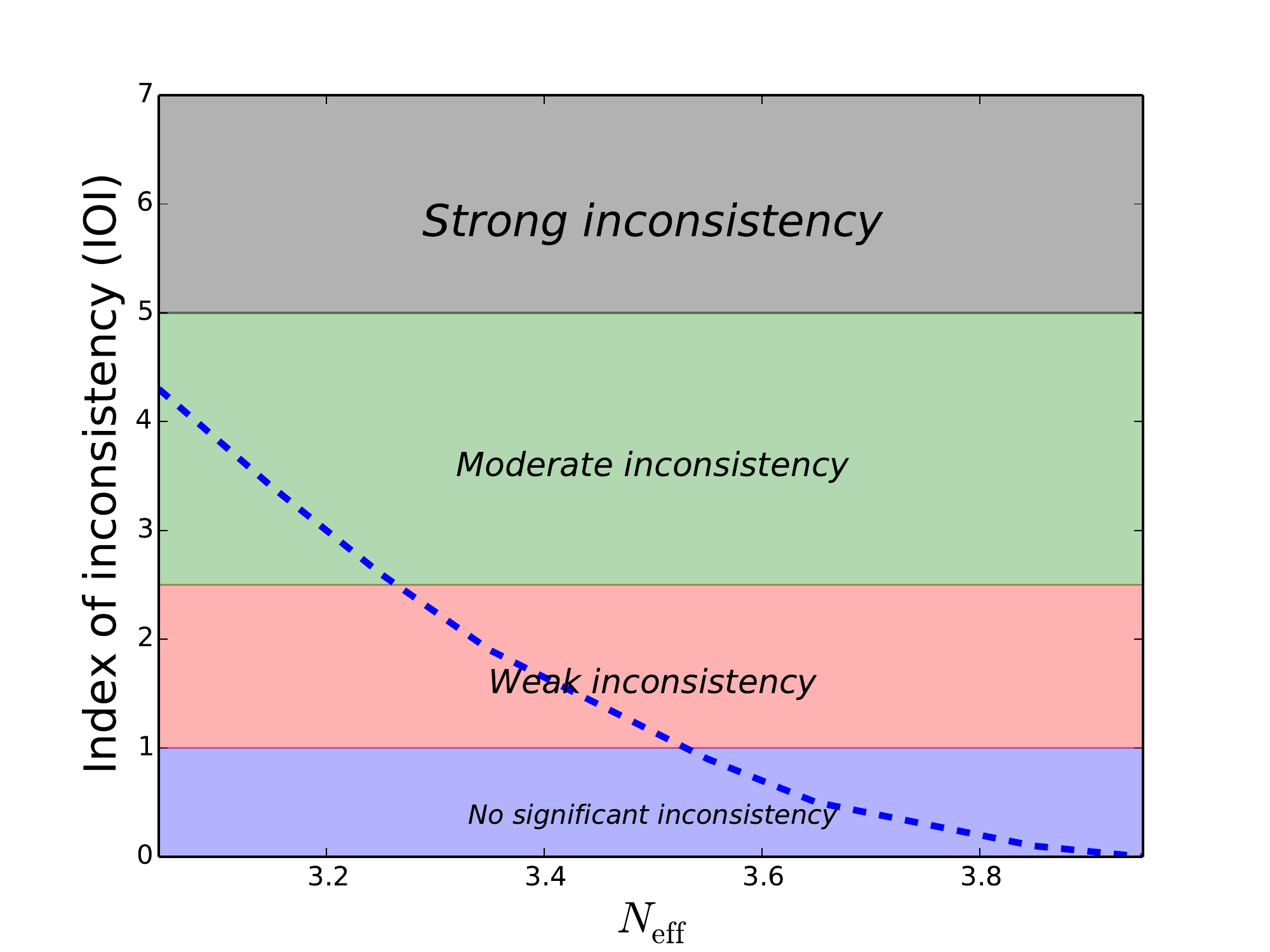}
\caption{As in Fig.~\ref{fig:ioi_w} but for the $\overline{N}\Lambda$CDM model.}
\label{fig:ioi_neff}
\end{figure}

The results discussed above are visually summarized in Fig.~\ref{fig:H0_neff_tension_and_evidence} and Fig.~\ref{fig:ioi_neff} (completely analogous to their counterparts for the $\overline{w}$CDM model, Fig.~\ref{fig:H0_w_tension_and_evidence} and Fig.~\ref{fig:ioi_w}). Fig.~\ref{fig:H0_neff_tension_and_evidence} shows how the tension measured in $\#\sigma$ rapidly decreases as $N_{\rm eff}$ moves towards $3.95$, at the cost however of adopting a model which is disfavoured with respect to $\Lambda$CDM (as quantified by the rapidly increasing value of $-\ln B_{ij}$, except within the region $3.046 \lesssim N_{\rm eff} \lesssim 3.25$ where the $\overline{N}\Lambda$CDM model is actually weakly favoured). In general, the results for the $\overline{N}\Lambda$CDM model are slightly more encouraging than for the $\overline{w}$CDM model. In fact, from the same figure we see that accepting a $\overline{N}\Lambda$CDM model which is weakly disfavoured with respect to $\Lambda$CDM (blue shaded region, $-\ln B_{ij}<1$), the tension can be brought almost to the $1.5\sigma$ level (which depending on personal taste might be enough for the $H_0$ tension to be considered solved), whereas accepting a $\overline{N}\Lambda$CDM model which is definitely disfavoured with respect to $\Lambda$CDM (pink shaded region, $-\ln B_{ij}<3$), the tension can be brought down to the $0.8\sigma$ level.

Analogously to what I did for the $\overline{w}$CDM model, I can estimate the dimensionless multiplier relating variations in $H_0$ to variations in the fixed value of $N_{\rm eff}$, which reflects the direction and strength of the $H_0$-$N_{\rm eff}$ correlation, which I will later show in Fig.~\ref{fig:basennu_tri}: $\Delta H_0 \approx m_N\Delta N_{\rm eff}$. I numerically estimate $m_N \approx 6.2$, and hence:
\begin{eqnarray}
\Delta H_0 = H_0-H_0\vert_{\Lambda{\rm CDM}} \approx 6.2\Delta N_{\rm eff}=6.2(N_{\rm eff}-3.046)\,, \nonumber \\\label{eq:mn}
\end{eqnarray}
where $H_0\vert_{\Lambda{\rm CDM}}$ is the value of $H_0$ inferred within $\Lambda$CDM. As with Eq.~(\ref{eq:mw}), Eq.~(\ref{eq:mn}) is useful in light of continuous updates in the local distance ladder measurement of $H_0$. For instance, using the latest value of $H_0$ reported in~\cite{Riess:2019cxk} and Eq.~(\ref{eq:mn}), I find that a physical theory would need to fix $N_{\rm eff} \approx 4.15$ in order to restore perfect agreement between the high-redshift and local measurements of $H_0$. Again, the caveat is that the coefficient $6.2$ should be updated if future CMB, BAO, or SNe datasets are used. As previously, one should keep in mind that the dimensionless multipliers only account for shifts in the central values of the $H_0$ posterior, but do not account for the fact that uncertainties in the local value of $H_0$ are continuously shrinking.

In Fig.~\ref{fig:ioi_w}, I plot the index of inconsistency as a function of the fixed value of $N_{\rm eff}$ in the $\overline{N}\Lambda$CDM class of models. One sees that for $3.046 \lesssim N_{\rm eff} \lesssim 3.25$, the inconsistency between the two measurements is moderate, whereas the inconsistency becomes weak for $3.25 \lesssim N_{\rm eff} \lesssim 3.55$ and insignificant for $3.55 \lesssim N_{\rm eff} \lesssim 3.95$.

In conclusion, in this part of the work I have examined the possibility of addressing the $H_0$ tension assuming that a physical theory is able to fix (or approximately fix) $w$ and $N_{\rm eff}$ to non-standard values within the $\overline{w}$CDM and $\overline{N}\Lambda$CDM models. I have found that it is not possible to completely remove the tension (\textit{i.e.} obtain a high-redshift estimate of $H_0$ that is in perfect agreement with the local distance ladder estimate) without incurring in a model which is strongly disfavoured against $\Lambda$CDM from a Bayesian evidence standpoint. In particular, the $H_0$ tension is completely removed if a physical theory is able to fix $w=-1.3$ [$H_0 = (73.2 \pm 0.7)\,{\rm km}\,{\rm s}^{-1}\,{\rm Mpc}^{-1}$] or $N_{\rm eff}=3.95$ [$H_0 = (73.1 \pm 0.6)\,{\rm km}\,{\rm s}^{-1}\,{\rm Mpc}^{-1}$], leading however to two models which are both strongly disfavoured from the Bayesian evidence standpoint with respect to $\Lambda$CDM ($\ln B_{ij} = -14.9$ and $\ln B_{ij} = -5.5$ respectively).

\subsection{Extended models}
\label{subsec:extended}

How does my non-standard approach adopted so far compare to the more standard approach where extended models with additional free parameters are considered? Notice that within the standard approach typically a prior on $H_0$ consistent with the local distance ladder measurement is also added to the standard high-redshift data. This contributes to ``pushing'' $H_0$ towards higher values, further reducing the $H_0$ tension. However, it is not always clear whether including such a prior is a consistent and legitimate operation to begin with.

To address this question, I consider the three extended models described in Sec.~\ref{sec:data}: $\Lambda$CDM+$w$, $\Lambda$CDM+$N_{\rm eff}$, and $\Lambda$CDM+$w$+$N_{\rm eff}$. I estimate $H_0$ within these three models by combining the \textit{cosmo} CMB+BAO+SNe dataset with a prior on $H_0$ consistent with the local distance ladder measurement of~\cite{Riess:2016jrr}. I compare the value inferred for $H_0$ to its local distance ladder value, and assess the statistical preference (if any) for these extended models against $\Lambda$CDM by computing their Bayesian evidence.~\footnote{Note that a fair comparison with $\Lambda$CDM should be made using the same datasets. In other words when computing $\ln B_{ij}$ using \texttt{MCEvidence}, the $\Lambda$CDM MCMC chains I utilize are obtained combining the \textit{cosmo} CMB+BAO+SNe dataset with the same prior on $H_0$.}

\begin{figure}[!tbh]
\includegraphics[width=1.0\linewidth]{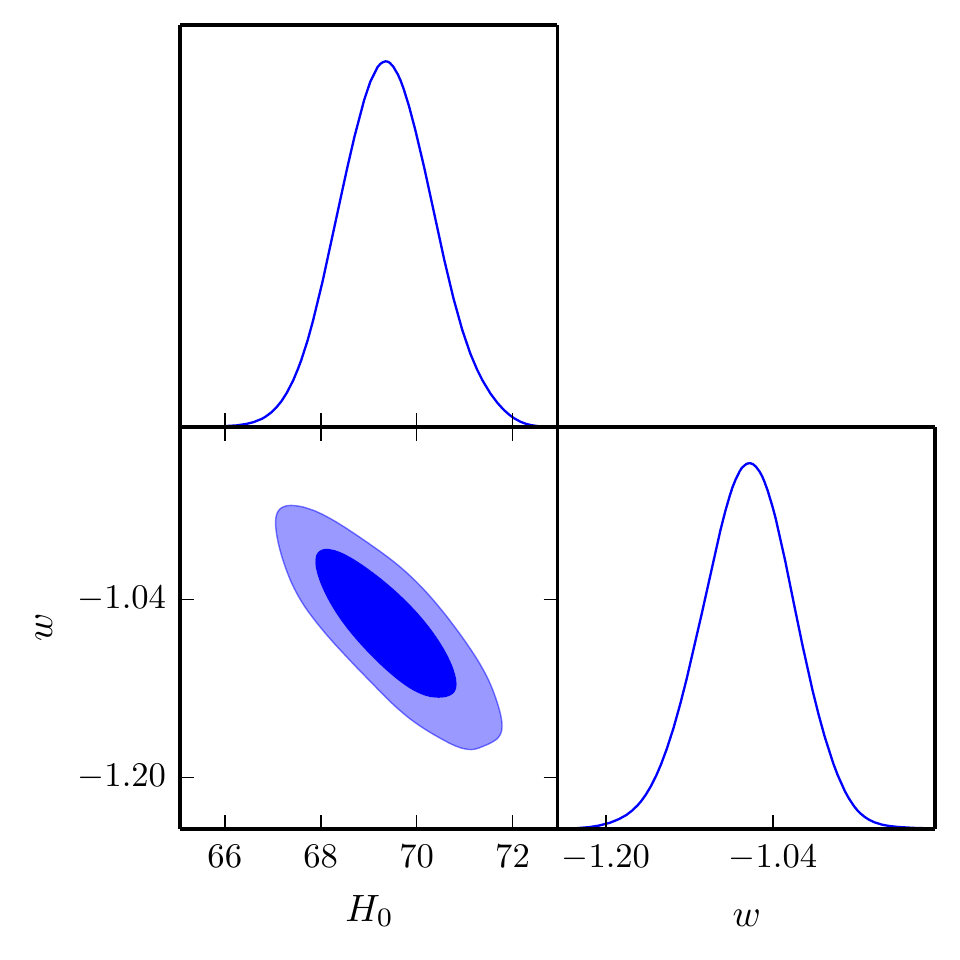}
\caption{Triangular plot showing the 2D joint and 1D marginalized posterior distributions for the Hubble constant $H_0$ and the dark energy equation of state $w$, obtained within the $\Lambda$CDM+$w$ model (see Tab.~\ref{tab:models}) and combining the \textit{cosmo} CMB+BAO+SNe dataset with a Gaussian prior on $H_0 = (73.24 \pm 1.74)\,{\rm km}\,{\rm s}^{-1}\,{\rm Mpc}^{-1}$ consistent with the local distance ladder measurement. The plot clearly shows the strong anti-correlation between $H_0$ and $w$ (see Sec.~\ref{sec:theory} for further discussions), which explains why adding the local prior on $H_0$ pushes $w$ into the phantom ($w<-1$) regime.}
\label{fig:basew_tri}
\end{figure}

I begin by considering the one-parameter $\Lambda$CDM+$w$ extension where I allow the dark energy equation of state $w$ to vary freely. Considering the \textit{cosmo} CMB+BAO+SNe dataset in combination with a prior on $H_0$ based on the local distance ladder measurement, I infer $H_0 = (69.4 \pm 1.0)\,{\rm km}\,{\rm s}^{-1}\,{\rm Mpc}^{-1}$. On the other hand, I find a value of the dark energy equation of state of $w=-1.06\pm 0.04$, which lies in the phantom regime at $>1\sigma$. This is expected, given that the prior on $H_0$ tends to ``pull'' $w$ within the phantom regime, due to the strong anti-correlation between $H_0$ and $w$ I extensively discussed in Sec.~\ref{sec:theory}. In Fig.~\ref{fig:basew_tri} I show the 2D joint and 1D marginalized posterior distributions of $H_0$ and $w$, which clearly show the strong anti-correlation between the two parameters.

We see that within the $\Lambda$CDM+$w$ model, the tension with the local measurement of $H_0$ is reduced to the level of $\approx 1.9 \sigma$, but not completely removed. As anticipated earlier, this reduction is partially attributable to the increase in error bar due to the extended parameter space (\textit{i.e.} marginalizing over the extra parameter $w$), and not to a genuine shift in the central value of the posterior of $H_0$ (as in the case of the $\overline{w}$CDM model when $w=-1.3$), which has only moved up to $H_0=69.4\,{\rm km}\,{\rm s}^{-1}\,{\rm Mpc}^{-1}$. When comparing this model against $\Lambda$CDM, I find that $\ln B_{ij}=-5.3$, corresponding to a very strong preference for $\Lambda$CDM.

\begin{figure}[!tbh]
\includegraphics[width=1.0\linewidth]{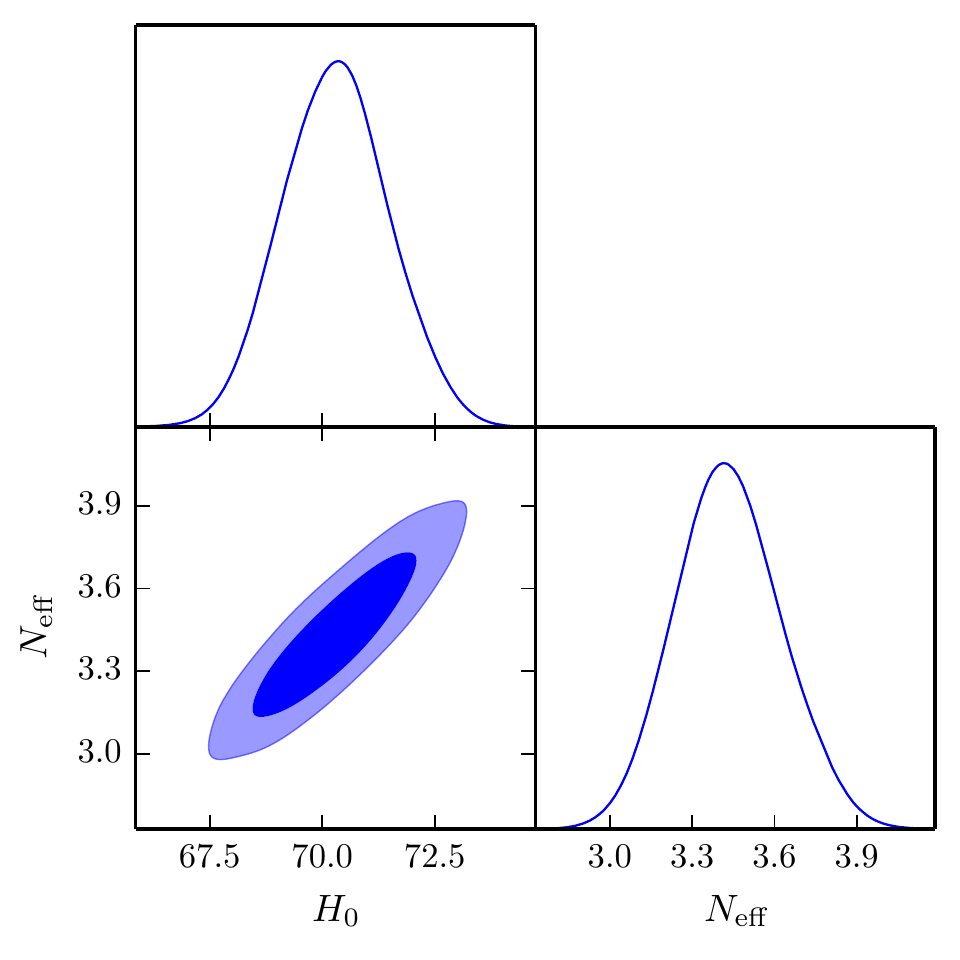}
\caption{Triangular plot showing the 2D joint and 1D marginalized posterior distributions for the Hubble constant $H_0$ and the effective number of relativistic species $N_{\rm eff}$, obtained within the $\Lambda$CDM+$N_{\rm eff}$ model (see Tab.~\ref{tab:models}) and combining the \textit{cosmo} CMB+BAO+SNe dataset with a Gaussian prior on $H_0 = (73.24 \pm 1.74)\,{\rm km}\,{\rm s}^{-1}\,{\rm Mpc}^{-1}$ consistent with the local distance ladder measurement. The plot clearly shows the strong correlation between $H_0$ and $N_{\rm eff}$ (see Sec.~\ref{sec:theory} for further discussions), which explains why adding the local prior on $H_0$ leads to a detection of extra relativistic species ($N_{\rm eff}>3.046$).}
\label{fig:basennu_tri}
\end{figure}

I now repeat this analysis within the $\Lambda$CDM+$N_{\rm eff}$ model, where I allow the effective number of relativistic species $N_{\rm eff}$ to vary freely. Combining the \textit{cosmo} CMB+BAO+SNe dataset with the local $H_0$ prior, I find $H_0 = (70.3 \pm 1.2)\,{\rm km}\,{\rm s}^{-1}\,{\rm Mpc}^{-1}$ and $N_{\rm eff}=3.43\pm 0.19$, which corresponds to a $\approx 2\sigma$ detection of extra relativistic species, again expected given the strong correlation between $H_0$ and $N_{\rm eff}$ discussed in Sec.~\ref{sec:theory} (see also the triangular plot in Fig.~\ref{fig:basennu_tri}).

As in the $\Lambda$CDM+$w$ case, the tension with the local distance ladder measurement of $H_0$ is reduced (this time to the level of $\approx 1.4 \sigma$), but not completely removed, and this is again partially attributable to the increase in error bar due to the extended parameter space. Moreover, Bayesian evidence considerations again disfavour the $\Lambda$CDM+$N_{\rm eff}$ model with respect to $\Lambda$CDM. In fact, I find $\ln B_{ij}=-4.6$, corresponding to a strong preference for $\Lambda$CDM.

\begin{figure}[!tbh]
\includegraphics[width=1.0\linewidth]{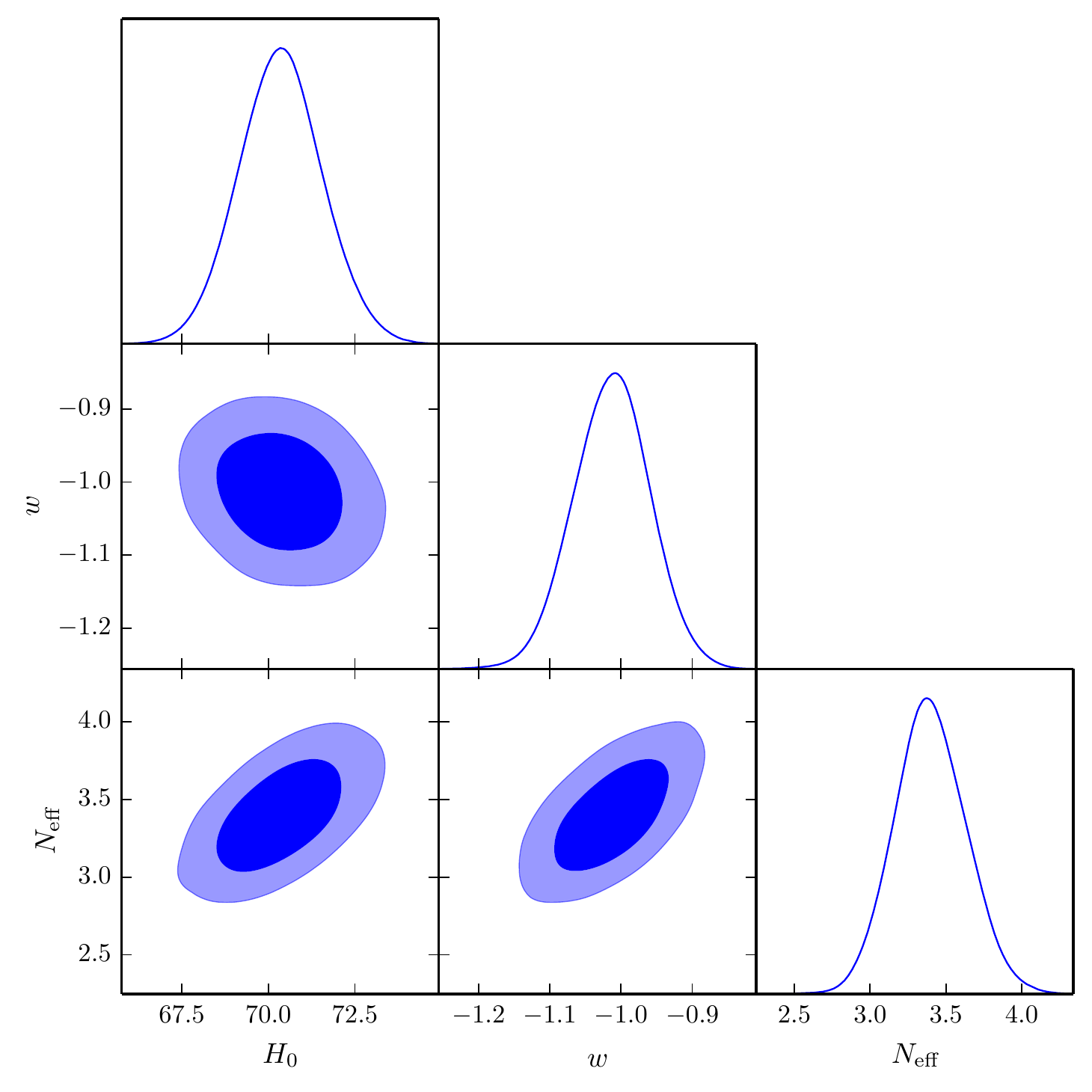}
\caption{Triangular plot showing the 2D joint and 1D marginalized posterior distributions for the Hubble constant $H_0$, the dark energy equation of state $w$, and the effective number of relativistic species $N_{\rm eff}$, obtained within the $\Lambda$CDM+$w$+$N_{\rm eff}$ model (see Tab.~\ref{tab:models}) and combining the \textit{cosmo} CMB+BAO+SNe dataset with a Gaussian prior on $H_0 = (73.24 \pm 1.74)\,{\rm km}\,{\rm s}^{-1}\,{\rm Mpc}^{-1}$ consistent with the local distance ladder measurement. The plot clearly shows the strong anti-correlation between $H_0$ and $w$, and the strong correlation between $H_0$ and $N_{\rm eff}$ (see Sec.~\ref{sec:theory} for further discussions).}
\label{fig:basewnnu_tri}
\end{figure}

I finally consider the two-parameter extension $\Lambda$CDM+$w$+$N_{\rm eff}$, where I allow both the dark energy equation of state $w$ and the effective number of relativistic species $N_{\rm eff}$ to freely vary. Combining the \textit{cosmo} CMB+BAO+SNe dataset with the local prior on $H_0$, I find $H_0=(70.3 \pm 1.2)\,{\rm km}\,{\rm s}^{-1}\,{\rm Mpc}^{-1}$, $w=-1.01 \pm 0.05$, and $N_{\rm eff}=3.40 \pm 0.24$. This time, both $w$ and $N_{\rm eff}$ are consistent with their standard values of $w=-1$ and $N_{\rm eff}$ within $2\sigma$ (see also the triangular plot in Fig.~\ref{fig:basewnnu_tri}).

Within the $\Lambda$CDM+$w$+$N_{\rm eff}$ model, the tension with the local measurements of $H_0$ is reduced to the level of $\approx 1.4\sigma$, but once more the reduction is partially attributable to the increase in error bar due to marginalization over two extra parameters. Bayesian evidence considerations also disfavour this model with respect to the baseline $\Lambda$CDM model. In fact, I find $\ln B_{ij}=-6.5$, which indicates a very strong preference for $\Lambda$CDM.

\subsection{Discussion}
\label{subsec:discussion}

\begin{figure}[!tbh]
\centering
\includegraphics[width=1.0\linewidth]{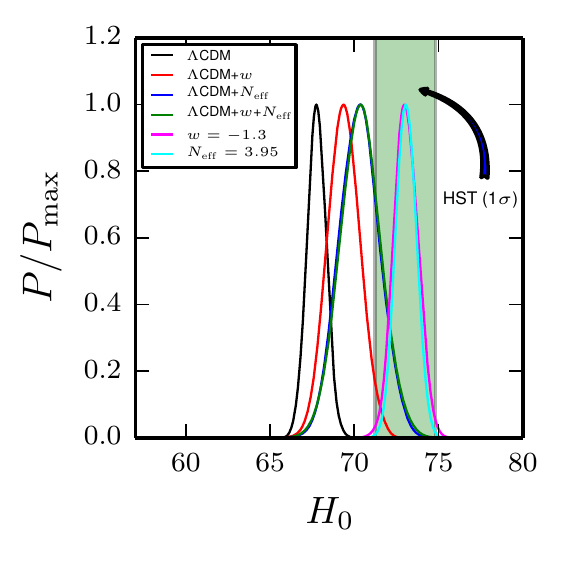}
\caption{Normalized posterior distributions of $H_0$, for a selection of models discussed in the text: the baseline $\Lambda$CDM model where $w=-1$ and $N_{\rm eff}=3.046$ (black curve), the $\overline{w}$CDM model with $w$ fixed to $w=-1.3$ (magenta curve), the $\overline{N}\Lambda$CDM model with $N_{\rm eff}$ fixed to $N_{\rm eff}=3.95$ (cyan curve), the $\Lambda$CDM+$w$ one-parameter extension of $\Lambda$CDM where $w$ is free to vary (red curve), the $\Lambda$CDM+$N_{\rm eff}$ one-parameter extension of $\Lambda$CDM where $N_{\rm eff}$ is free to vary (blue curve), and the $\Lambda$CDM+$w$+$N_{\rm eff}$ two-parameter extension of $\Lambda$CDM where both $w$ and $N_{\rm eff}$ are free to vary (green curve). The green shaded region is the 1$\sigma$ credible region for $H_0$ determined by the local distance ladder measurement of \textit{HST}~\cite{Riess:2016jrr}, yielding $H_0 = (73.24 \pm 1.74)\,{\rm km}\,{\rm s}^{-1}\,{\rm Mpc}^{-1}$. The former three posteriors are obtained using the \textit{cosmo} CMB+BAO+SNe dataset combination, whereas the latter three include in addition a Gaussian prior on $H_0$ based on the local distance ladder measurement, which further helps pulling the posteriors towards higher values of $H_0$. Within the $\overline{w}$CDM model with $w=-1.3$ and the $\overline{N}\Lambda$CDM model with $N_{\rm eff}=3.95$, the high-redshift estimateis in complete agreement with the local distance ladder measurement (\textit{i.e.} the tension is brought down to $0\sigma$), whereas within the three extended models the tension is only partially addressed, and partly due to an increase in the uncertainty.}
\label{fig:H0_posterior_extended}
\end{figure}

I will now provide a critical discussion of the results obtained in Sec.~\ref{subsec:nonstandard} and Sec.~\ref{subsec:extended}, comparing the two approaches towards addressing the $H_0$ tension: assuming that a physical theory is able to fix $w$ and $N_{\rm eff}$ to non-standard values versus considering extended models. A visual comparison of the posterior distributions for 6 representative cases discussed earlier is presented in Fig.~\ref{fig:H0_posterior_extended}, alongside the $1\sigma$ region for $H_0$ based on the local distance ladder measurement. As is very clear from the figure, the three extended models considered in Sec.~\ref{subsec:extended} only partially address the tension, mostly through a broadening of the posterior distribution due to marginalization over 1 or 2 additional parameters (and partially helped by including a prior on $H_0$ based on the local distance ladder value, which contributes to ``pulling'' the value of $H_0$ up). On the other hand, the $\overline{w}$CDM model with $w=-1.3$ and the $\overline{N}\Lambda$CDM model with $N_{\rm eff}=3.95$ genuinely address the tension by shifting the posterior distribution to overlap with the local distance ladder measurements. However, for both the extended and non-standard models, all of this comes at the price of considering models which are strongly disfavoured against $\Lambda$CDM from the Bayesian evidence point of view.

Within the $\overline{w}$CDM and $\overline{N}\Lambda$CDM models, it is not possible to lower the $H_0$ tension at a level $\lesssim 1\sigma$ while at the same time dealing with a model which is not strongly disfavoured against $\Lambda$CDM from a Bayesian evidence point of view. My analysis also reveals that the situation is somewhat less dramatic if a physical theory were able to fix $N_{\rm eff}$ rather than $w$ to non-standard values (see, for instance, the difference $\Delta \ln B_{ij} \approx 10$ between the values of $\ln B_{ij}$ obtained when $w=-1.3$ versus $N_{\rm eff}=3.95$, and discussed above). The reason is that low-redshift measurements of the expansion history (BAO and SNe) exquisitely constrain the dark energy equation of state to be very close to that of a cosmological constant, leaving very little freedom in modifying the late-time dynamics and in particular the equation of state of dark energy without incurring into a bad fit to the data. On the other hand, there is significantly more freedom available in modifying the early expansion history through $N_{\rm eff}$ which is unconstrained by low-redshift data (recall that in addition I have made the conservative choice of not including small-scale polarization data, which would help constraining $N_{\rm eff}$ but could still be contaminated by systematics).

Concerning the 3 extended models I considered ($\Lambda$CDM+$w$, $\Lambda$CDM+$N_{\rm eff}$, and $\Lambda$CDM+$w$+$N_{\rm eff}$), one sees that all 3 are strongly/very strongly disfavoured with respect to $\Lambda$CDM, yielding $\ln B_{ij}=-5.3$, $\ln B_{ij}=-4.6$, and $\ln B_{ij}=-6.5$ respectively. Aside from not being able to satisfactorily solve the $H_0$ tension, the three models are penalized by the presence of extra parameters, which are not justified by the improvement in fit.

One interesting point of discussion could then be the following. Let us assume we are willing to tolerate a certain amount of residual tension between the two measurements of $H_0$. For instance, most of the works aiming to address the $H_0$ tension consider the tension solved if it drops below the $1.5\sigma-2\sigma$ level. In this work, my initial aim was to determine what values of $w$ and $N_{\rm eff}$ a physical theory should be able to predict to bring the tension down to essentially $0\sigma$, which is perhaps rather ambitious! Let me instead be more open and choose $2\sigma$ as a threshold for considering the tension solved to a satisfactory extent. Then, from Fig.~\ref{fig:H0_w_tension_and_evidence} and Fig.~\ref{fig:H0_posterior_combined_neff}, we see that this can be achieved at the expense of considering models where $w \approx -1.07$ or, better still, $N_{\rm eff} \approx 3.5$, which are ``only'' weakly disfavoured with respect to $\Lambda$CDM ($-\ln B_{ij}<1$). The $2\sigma$ threshold is of course a subjective threshold, and I have introduced it simply for the sake of argument. The take away message is that even if a theory were able to fix $w$ and $N_{\rm eff}$ to non-standard values which are not strongly disfavoured from the Bayesian evidence point of view, this might be sufficient to lower the $H_0$ tension to a level where the tension might be considered at least partially addressed.

In fact, let me take one step forward and compare the $\overline{N}\Lambda$CDM model with the extended $\Lambda$CDM+$N_{\rm eff}$ model. Of the three extended models, the latter was able to reduce the tension the most (down to the $1.4\sigma$ level), while at the same time being least disfavoured from the Bayesian evidence point of view (albeit leading to $\ln B_{ij}=4.6$ and still being strongly disfavoured with respect to $\Lambda$CDM). The question then is: what is the price to pay to construct a $\overline{N}\Lambda$CDM model which fares as well (or better) than $\Lambda$CDM+$N_{\rm eff}$? In other words, what is the minimum $-\ln B_{ij}$ for a $\overline{N}\Lambda$CDM model which reduces the $H_0$ tension below the $1.4\sigma$ level? We immediately read off the answer from Fig.~\ref{fig:H0_neff_tension_and_evidence}: a minimum $-\ln B_{ij}$ of $\approx 1.3$, which is obtained by considering $N_{\rm eff} \approx 3.55$, is required to lower the $H_0$ tension below $1.4\sigma$ within the $\overline{N}\Lambda$CDM model. This is somewhat surprising and interesting: we have found a (class of) non-standard models which performs equally well in terms of lowering the statistical significance of the $H_0$ tension compared to a similar extended model, but is less disfavoured from the Bayesian evidence point of view with respect to $\Lambda$CDM (the $\overline{N}\Lambda$CDM model with $N_{\rm eff}=3.55$ is ``only'' definitely disfavoured with $-\ln B_{ij}=1.3$, as opposed to the $\Lambda$CDM+$N_{\rm eff}$ model which is strongly disfavoured with $-\ln B_{ij}=4.6$).

I can repeat the same exercise for the $\overline{w}$CDM model, which as argued earlier faces more difficulties compared to the $\overline{N}\Lambda$CDM model since low-redshift data tends to favour a value for $w$ very close to the standard $-1$. This makes it really difficult to lower $w$ significantly into the phantom regime without incurring into a very low value of the Bayesian evidence. I address the same question as earlier: what is the minimum $-\ln B_{ij}$ for a $\overline{w}$CDM model which reduces the $H_0$ tension below the $1.4\sigma$ level? Again, we can read off the answer from Fig.~\ref{fig:H0_w_tension_and_evidence}: a minimum $-\ln B_{ij}$ of $\approx 4$, which is obtained by considering $w \approx -1.15$, is required to lower the $H_0$ tension below $1.4\sigma$ within the $\overline{w}\Lambda$CDM model. Again, this is a very surprising result: despite the difficulties, the $\overline{w}$CDM model with $w=-1.15$ and $-\ln B_{ij}=4.0$ still performs better than the $\Lambda$CDM+$N_{\rm eff}$ model (for which, recall, $-\ln B_{ij}=4.6$) from the Bayesian evidence point of view, while lowering the $H_0$ tension down to the same level of significance.

What is the take away message from these two exercises? All things being equal (\textit{i.e.} the $H_0$ tension being lowered to the same statistical significance, which I took to be $1.4\sigma$ in the above example, or $2\sigma$ earlier), it is more efficient from the Bayesian evidence point of view to consider physical theories which are able to fix $N_{\rm eff}$ and $w$ to non-standard values (for $N_{\rm eff}=3.95$ and $w=-1.3$ I obtained $\ln B_{ij}=-1.3$ and $\ln B_{ij}=-4.0$ respectively, as opposed to $\ln B_{ij}=-4.6$ for $\Lambda$CDM+$N_{\rm eff}$). In addition, these non-standard models lower the $H_0$ tension by actually shifting the posterior distribution without broadening it, leading to a somewhat more appealing resolution. On the other hand, when choosing between a physical theory able to fix $w$ or $N_{\rm eff}$ to non-standard values, my analysis reveals that the latter is preferable.

Finally, it is worth remarking that the comparison I have made above is also somewhat penalizing the non-standard models compared to the extended ones. In fact, when inferring $H_0$ within the extended models I have also included a prior on $H_0$ based on the local distance ladder measurement, which of course helps raising $H_0$ towards the local value. On the other hand, such a prior was not included when inferring $H_0$ within the non-standard $\overline{w}$CDM and $\overline{N}\Lambda$CDM models: including it would only strengthen the conclusion I reached above.

\subsection{Models predicting fixed values for the extra parameters}
\label{subsec:models}

So far, I have discussed the $H_0$ tension in light of possible models which would allegedly be able to fix extra beyond-$\Lambda$CDM parameters (such as $w$ or $N_{\rm eff}$) to non-standard values. I have found that models fixing the effective number of relativistic species to $N_{\rm eff} \approx 3.95$ or the dark energy equation of state to $w \approx -1.3$ can completely remove the $H_0$ tension at the cost of a worsened fit to CMB, BAO, and SNeIa data, whereas less extreme values can improve the fit while still reducing the $H_0$ tension considerably. Throughout this discussion, however, an elephant in the room in the form of the following question remains: ``\textit{Do such models exist in first place?}'' In general, most models will predict a range of values for the extra parameters, whose precise value will depend on specific theory parameters (such as the values of the Lagrangian couplings, or the specific form of the kinetic term or potential of a dark energy field). On the understanding that the existence or not of such models does not undermine the motivation for the present work (which should rather be seen as providing model builders with parameter values to test against), in the following I will briefly discuss a number of theoretical models which are able to fix, or approximately fix, $w$ and $N_{\rm eff}$ near their ``sweet spot'' values. The existence of such models further reinforces the motivation behind this work, adding value to the proposed exercise and making the exercise itself more compelling.

I begin by discussing theoretical models which are able to approximately fix $w \approx -1.3$. An example of one such model is the vector-like dark energy model constructed in~\cite{ArmendarizPicon:2004pm}. This model is constructed out of a ``cosmic triad", \textit{i.e.} a set of three identical one-forms pointing in mutually orthogonal spatial directions, in such a way as to respect isotropy. Another field-based model of phantom dark energy predicting $w \approx -1.29$ is the phantom Dirac-Born-Infeld model constructed in~\cite{Barenboim:2017sjk}, with Hamiltonian bounded from below in the comoving frame (although not in every frame). Other works have argued that phantom dark energy models could naturally arise from string theory, due to the correlation between winding and momentum modes in conjunction with an exponentially falling angular frequency. An example is~\cite{Frampton:2002tu}, where a concrete string theory model predicting $w \approx -4/3$ is constructed.

Modifications of General Relativity also provide a route towards constructing stable effective phantom components. In this context, in~\cite{Nojiri:2009pf} it was argued that a coupled phantom model where dark matter is coupled to a phantom dark energy component with $w=-4/3$ could cure the coincidence problem, behave as an attractor at late times, and avoid the Big Rip singularity. In~\cite{Nojiri:2005sx}, a phantom DE model with finite-time future singularity not of the Big Rip type, where at late times the DE EoS $w=-4/3$ behaves as a stable fixed point (attractor), is constructed. The type of singularity achieved in this model, as well as the model itself, could be motivated by the finite action principle proposed by Barrow and Tipler~\cite{Barrow:1988ghw,Barrow:2019gzc}.

Rather than arising from a fundamental action principle (either in the context of additional fields or modifications to General Relativity), models with $w \approx -1.3$ could have a more profound symmetry-based motivation, or mimic something else altogether. An example is given in~\cite{Dabrowski:2003jm,Dabrowski:2007dp} in terms of the so-called \textit{phantom duality}, a symmetry mapping models with equation of state $w \to -(2+w)$. This duality implies that domain walls (well-motivated topological defects) whose effective EoS is $w=-2/3$ are dual to phantom models with $w=-4/3 \approx -1.3$. In~\cite{Dabrowski:2003jm,Dabrowski:2007dp}, the phantom duality is argued to be quite fundamental and closely related to the scale factor duality in pre-big-bang models, itself motivated by superstring cosmology scale factor duality symmetries. The phantom duality provides a fundamental motivation for considering phantom models with $w=-4/3$. Returning to the possibility of models with $w \approx -1.3$ mimicking something else altogether, a possibility in this sense is presented in~\cite{Godlowski:2004pt}. There it is argued that a component with $w=-4/3$ would naturally arise in extra dimensional models such as the Randall-Sundrum model. This component would mimic $\Lambda_{(4)}$, the 4-dimensional cosmological constant induced by the projection of the 5-dimensional Randall-Sundrum Friedmann equations on the brane, where this component with $w=-4/3$ would reside.

Finally, the so-called quantum bias model for dark energy~\cite{Butcher:2017asw,Butcher:2018mth}, where the time-dependent information capacity in discarded degrees of freedom could drive cosmic acceleration, generically predicts a phantom dark energy component. The model does not unambiguously predict a value for $w$ at present time as the latter depends on the free parameter $\bar{d}$. However, for $\bar{d} \approx 3$ one recovers $w \approx -1.3$, where $\bar{d} \approx 3$ could be strongly motivated from first principles given that we appear to live in 3 spatial dimensions.~\footnote{From private communication with the author Luke Butcher.}

So far I have discussed models which are able to fix $w$. Let me now discuss models which are able to fix, or approximately fix, $N_{\rm eff}$: one could subjectively argue that such models are less exotic than the dark energy models I discussed above. A value of $N_{\rm eff} \approx 4$ indicates at face value an almost fully thermalized extra relativistic species. One interesting possibility in this sense is the possibility of a fully thermalized sterile neutrino. This could be motivated by a series of short-baseline anomalies in reactor neutrino experiments, among which the MiniBooNE anomaly~\cite{Aguilar-Arevalo:2018gpe}. In fact, the best-fit mass-squared splitting and mixing angle for a sterile neutrino explanation of the MiniBooNE anomaly lead to almost complete thermalization (\textit{i.e.} $N_{\rm eff} \approx 4$), as shown in e.g.~\cite{Gariazzo:2015rra,Giunti:2019aiy,Gariazzo:2019gyi,Diaz:2019fwt,Boser:2019rta,Adams:2020nue,Hagstotz:2020ukm}.

Moving to other models predicting less extreme values of $N_{\rm eff}$, a single thermally decoupled pseudo-Nambu-Goldstone boson (pNGB) can lead to rather specific predictions for $N_{\rm eff}$ depending on the temperature at which the pNGB freezes out (see e.g. Fig.~1 of~\cite{Baumann:2016wac}). For example, freeze-out occurring just after the QCD phase transition would predict $N_{\rm eff} \approx 3.4$, whereas freeze-out occurring between $100\,{\rm MeV}$ and $1\,{\rm MeV}$ would predict $N_{\rm eff} \approx 3.7$. As an additional example, the model studied by Weinberg in~\cite{Weinberg:2013kea}, featuring an extra Goldstone boson possibly associated to a dark matter particle number $U(1)'$ symmetry, predicts $N_{\rm eff} \approx 3.45$.

As I discussed earlier, models with extra Abelian symmetries generally predict a higher value of $N_{\rm eff}$. One example of such model which also predicts a rather specific value of $N_{\rm eff}$ is the abelian $L_{\mu}-L_{\tau}$ extension of the Standard Model studied in~\cite{Escudero:2019gzq}, which predicts $N_{\rm eff} \approx 3.25$ across most of its parameter space.~\footnote{Note, however, that this model does not produce extra relativistic species in the usual sense, but rather injects extra energy to the Standard Model neutrinos through the decay of a light and weakly coupled Z' vector boson.} The Majoron, a light weakly coupled neutrino-philic scalar associated to the spontaneous breaking of lepton number symmetry, also leads to very specific predictions for $N_{\rm eff}$. For example, a single Majoron associated to a Dirac neutrino mass generation mechanism predicts $N_{\rm eff}=3.15$ across a wide range of parameter space as shown for instance in~\cite{Escudero:2019gvw}, while $N_{\rm eff}=3.35$ if the neutrino mass generation mechanism is Majorana. Allowing for more than 1 Majoron and the neutrino mass generation being either Dirac or Majorana, the predictions for $N_{\rm eff}$ could lie anywhere between $3.15$ and $4.05$, with the specific value depending on the number of Majorons and mass generation mechanism (see e.g. Tab.~3 of~\cite{Chacko:2003dt}). However, it is important to stress that once these two are fixed (as a well-motivated theory does), the value of $N_{\rm eff}$ is a \textit{prediction}, \textit{i.e.} it does not vary as a function of other parameters.

Finally, turning to other models predicting Abelian extensions of the Standard Model, mirror dark matter with kinetic mixing parameter $\epsilon \sim 3 \times 10^{-9}$ (with this specific value motivated by solving the small-scale structure problems of collisionless cold dark matter while explaining galactic scaling relations and being consistent with null results from direct detection experiments~\cite{Foot:2016wvj}) predicts $N_{\rm eff} \approx 3.55$, as shown in~\cite{Foot:2011ve}. This was also shown in a more general setting in~\cite{Foot:2014uba}.

\section{Conclusions}
\label{sec:conclusions}

The persisting $H_0$ tension might be an indication for new physics beyond the concordance $\Lambda$CDM model. Most of the solutions considered so far (many of which invoking new physics in the dark sector of the Universe) involve \textit{extended} cosmological models, \textit{i.e.} extensions of the baseline $\Lambda$CDM model where additional parameters are allowed to vary. Two rather economical solutions in this direction involve either a phantom dark energy component (\textit{i.e.} a dark energy component with equation of state $w$ satisfying $w<-1$) or extra relativistic species in the early Universe (\textit{i.e.} $N_{\rm eff}>3.046$). Importantly, in these and several other extended models, the $H_0$ tension is only partially relieved, partly thanks to a broadening of the $H_0$ posterior distribution due to marginalization over additional free parameters rather than a genuine shift in the mean of the distribution itself (see e.g. Fig.~\ref{fig:H0_posterior_extended}).

In this work, I have considered an alternative approach. Focusing on the dark energy equation of state $w$ and the effective number of relativistic species $N_{\rm eff}$, I have asked the following questions: what value of $w$ or $N_{\rm eff}$ would a physical theory have to predict (so that the parameter itself can effectively be considered fixed) in order for the high-redshift estimate of $H_0$ from CMB, BAO, and SNeIa data to \textit{perfectly} match the local distance ladder estimate, \textit{i.e.} in order to formally reduce the $H_0$ tension to $\approx 0\sigma$? How much would Bayesian evidence considerations (dis)favour such a model with respect to $\Lambda$CDM? How does this approach compare, statistically speaking, to the standard one wherein the additional parameters are allowed to vary? Addressing these questions can prompt further model-building activity and provide model-builders with non-standard parameter values against which to test.

I have found (see Fig.~\ref{fig:H0_posterior_combined_w} and Fig.~\ref{fig:H0_posterior_combined_neff}) that a \textit{perfect} match between the high-redshift estimate of $H_0$ and the local distance ladder measurement (\textit{i.e.} reducing the tension to essentially $0\sigma$) can be achieved if a physical model is able to fix $w \approx -1.3$ or $N_{\rm eff} \approx 3.95$. Both are highly non-standard values for these parameters, and in fact Bayesian evidence considerations strongly disfavour the resulting non-standard models with respect to the baseline $\Lambda$CDM model ($\ln B_{ij}=-14.9$ and $\ln B_{ij}=-5.5$ respectively), see Fig.~\ref{fig:H0_w_tension_and_evidence} and Fig.~\ref{fig:H0_neff_tension_and_evidence}. I have then compared my approach to the more standard case where an attempt to address the $H_0$ tension is performed by allowing $w$ and/or $N_{\rm eff}$ to freely vary. Such extensions are able to lower the $H_0$ tension down to the $1.4-1.9\sigma$ level. However, they too are strongly disfavoured with respect to $\Lambda$CDM from Bayesian evidence considerations.

An interesting and somewhat more fair comparison is between extended ($w$ and $N_{\rm eff}$ varying) and non-standard models ($w$ and $N_{\rm eff}$ fixed to non-standard values) which reduce the $H_0$ tension to the same level of statistical significance (for instance, reducing the tension to the $1.5\sigma-2\sigma$ level will subjectively be considered by most to be a satisfying enough resolution to the $H_0$ tension). In this case I have found (see Sec.~\ref{subsec:discussion}) that perhaps somewhat surprisingly the non-standard models fare considerably better from the Bayesian evidence point of view. For instance, while the $\Lambda$CDM+$N_{\rm eff}$ extension is able to bring the tension down to the $1.4\sigma$ level at the expense of a value $\ln B_{ij}=-4.6$ strongly favouring $\Lambda$CDM, the tension can be brought down to the same level either if a physical model is able to fix $w \approx -1.15$ (which leads to $\ln B_{ij}=-4.0$) or even more efficiently if $N_{\rm eff} \approx 3.55$ (which leads to $\ln B_{ij}=-1.3$).

While the examples I have considered are limited, they appear to point to a perhaps unexpected fact: from the statistical point of view the $H_0$ tension does not seem to favour extensions to $\Lambda$CDM (a similar conclusion was already reached, through a different approach, in~\cite{Mortsell:2018mfj,Guo:2018ans}), but would rather prefer models which are able to fix (or approximately fix) the extra parameters to non-standard values. Such a conclusion can be particularly interesting for model-builders, with my results providing parameter values to test against. For instance, a well-motivated microphysical model making a definite prediction that $w \approx -1.3$ or $N_{\rm eff} \approx 3.95$ would also predict \textit{perfect} agreement between the high-redshift and local distance ladder estimates of $H_0$. Examples of models making predictions near the ``sweet spots'' considered are discussed in Sec.~\ref{subsec:models}. It is worth remarking, however, that neither of the approaches considered in this work has led to a fully satisfying resolution of the $H_0$ tension. None of the models considered here (be them extended or non-standard) are able to bring the tension below the $1\sigma$ level while not being excessively penalized by Bayesian evidence considerations. In this sense, a compelling solution to the $H_0$ tension with either approach remains to be found (see e.g.~\cite{Knox:2019rjx}).

One caveat of this work is that I have made use of the 2015 \textit{Planck} likelihood~\cite{Ade:2015xua} and compared results to the 2016 local distance ladder measurement of $H_0$~\cite{Riess:2016jrr}, whereas the new 2019 legacy \textit{Planck} likelihood was released in~\cite{Aghanim:2019ame} 2 weeks after this work appeared on arXiv, and new local distance ladder measurements of $H_0$ (which have increased the significance of the tension) are available~\cite{Riess:2019cxk}. Nonetheless, in Sec.~\ref{sec:results} [see particularly Eqs.~(\ref{eq:mw},\ref{eq:mn})] I have provided tools to estimate how much my results would change if one wished to adopt the same approach with a more updated value of $H_0$. In particular, I have numerically estimated dimensionless multipliers relating variations in $H_0$ to variations in $w$ and $N_{\rm eff}$. These suggest that my earlier results, valid for the $H_0$ measurement in~\cite{Riess:2016jrr}, would only slightly change when using the more updated measurement in~\cite{Riess:2019cxk}: in particular, one would need $w \approx -1.35$ and $N_{\rm eff} \approx 4.1$ in order to address the increased tension. I expect however that the earlier considerations comparing the statistical performance of the non-standard models against the corresponding extended models are robust to such changes. It would nonetheless be worth re-examining my approach when the new \textit{Planck} likelihood becomes available, or to perform forecasts in light of future CMB data, for instance from \textit{Simons Observatory}~\cite{Ade:2018sbj,Abitbol:2019nhf}.

In conclusion, in this work I have revisited the issue of addressing the $H_0$ tension by invoking new physics, adopting an alternative approach where I consider what would happen if a physical theory were able to fix a beyond-$\Lambda$CDM parameter to a specific value: in this case, the extra parameter is effectively fixed, and the model has the same number of parameters as $\Lambda$CDM. While the approach considered has not been able to address the $H_0$ tension in a statistically satisfactory way, I have demonstrated that from a purely statistical point of view the non-standard models considered fare as well, if not better, than their extended counterparts. The findings reported in this work might also have intriguing repercussions from the model-building perspective, providing model-builders non-standard values for the dark energy equation of state $w$ and the effective number of relativistic species $N_{\rm eff}$ to test against.

\section*{Acknowledgements}
It is a pleasure to thank Andrzej Drukier and Katherine Freese for raising a very interesting question which led me to develop the present work, Suhail Dhawan and Eleonora Di Valentino for many interesting discussions while this project was developed, Miguel Escudero for illuminating discussions regarding models able to fix $N_{\rm eff}$ to non-standard values, and the anonymous referee for very useful suggestions which helped me express my aims more clearly. I am supported by the Isaac Newton Trust and the Kavli Foundation through a Newton-Kavli fellowship, and acknowledge a College Research Associateship at Homerton College, University of Cambridge.

\bibliographystyle{JHEP}
\bibliography{hubble.bib}

\end{document}